\documentclass[journal]{IEEEtran}
\usepackage[utf8]{inputenc}

\usepackage{times}
\usepackage{amssymb}  
\usepackage{cite}     
\usepackage{textgreek}
\usepackage{upref}
\usepackage{amsfonts}
\usepackage[dvipsnames,usenames]{color}
\usepackage{cite}
\usepackage[inline]{enumitem}
\usepackage{caption}
\usepackage{balance}
\usepackage{nicefrac}
\usepackage{amsthm}
\usepackage[linesnumbered,ruled]{algorithm2e}
\usepackage{algorithmic}
\usepackage{verbatim}
\usepackage{times,color}
\usepackage{enumitem}
\usepackage{float}
\usepackage{color}
\usepackage{mathtools}
\usepackage{graphicx}
\usepackage{listings}
\usepackage{tikz}
\usepackage{multicol,multirow}
\usepackage{psfrag}
\usepackage{bbm}
\usepackage{soul}
\usepackage{varwidth}

\usepackage{pgfplots}
\usepackage{hhline}
\usepackage{xcolor,colortbl}

\usepackage{enumitem}

\usetikzlibrary{%
	arrows,%
	automata,%
	calc,%
	patterns,%
	plotmarks,%
	positioning,%
	shapes.geometric,%
	shapes,%
	snakes,%
	decorations.pathmorphing,%
	decorations.shapes,%
	graphs,%
	chains,%
	arrows.meta, %
}

\newcommand{\del}{\delta}

\newcommand{\be}[1]{\begin{equation}\label{#1}}
\newcommand{\ee}{\end{equation}}

\newcommand{\bc}{\begin{center}}
\newcommand{\ec}{\end{center}}
\newcommand{\lev}{\text{Lev}}

\newcommand{\cC}{{\cal C}}
\newcommand{\cD}{{\cal D}}

\newcommand{\cO}{{\mathcal O}}

\newcommand{\x}{\mathbf{x}}
\newcommand{\y}{\mathbf{y}}

\renewcommand{\leq}{\leqslant}

\renewcommand{\geq}{\geqslant}

\newcommand{\Cref}[1]{Co\-rol\-la\-ry\,\ref{#1}}

\newtheorem{theorem}{Theorem}

\newtheorem{definition}{Definition}

\newtheorem{corollary}[theorem]{Corollary} 
\newtheorem{remark}{Remark}
\newtheorem{claim}[theorem]{Claim}
\newtheorem{proposition}[theorem]{Proposition}

\newtheorem{example}{Example}

\newtheorem{construction}{Construction}

\definecolor{Codecolor}{named}{White}  

\newcommand{\Copen}{\mbox{\{\kern-5.50pt\{}}
\newcommand{\Cclose}{\mbox{\}\kern-5.50pt\}}}
\newcommand{\Cslash}{\mbox{$\backslash\kern-6.02pt\backslash$}}

\newcommand{\sepb}{{\color{green!50!black}0}}
\newcommand{\sepr}{{\color{red}1}}

\def \th{^\text{th}}

\begin{document}

\title{Optimal Codes Detecting Deletions in Concatenated Binary Strings Applied to Trace Reconstruction}%
\author{%
\IEEEauthorblockN{Serge~Kas Hanna\\ Department of Mathematics and Systems Analysis, School of Science, Aalto
University, Finland\\ Email: serge.kashanna@aalto.fi
}
\thanks{The results of this paper were presented in part at the 2021 and 2022 IEEE International Symposium on Information Theory~\cite{P1,P2}. }
}

\maketitle

\begin{abstract}
    Consider two or more strings $\x^1,\x^2,\ldots,$ that are concatenated to form $\x=\langle \x^1,\x^2,\ldots \rangle$. Suppose that up to $\del$ deletions occur in each of the concatenated strings. Since deletions alter the lengths of the strings, a fundamental question to ask is: how much redundancy do we need to introduce in $\x$ in order to recover the boundaries of $\x^1,\x^2,\ldots$? This boundary problem is equivalent to the problem of designing codes that can detect the exact number of deletions in each concatenated string. In this work, we answer the question above by first deriving converse results that give lower bounds on the redundancy of deletion-detecting codes. Then, we present a marker-based code construction whose redundancy is asymptotically optimal in $\del$ among all families of deletion-detecting codes, and exactly optimal among all block-by-block decodable codes. To exemplify the usefulness of such deletion-detecting codes, we apply our code to trace reconstruction and design an efficient coded reconstruction scheme that requires a constant number of traces.
\end{abstract}

\section{Introduction}
\label{sec:intro}
We study the problem of detecting deletions in concatenated binary strings. More precisely, we consider strings of the form \mbox{$\x=\langle \x^1, \x^2, \ldots, \x^{n/\ell} \rangle \in \mathbb{F}_2^n$}, that are implicitly divided into $n/\ell$ disjoint substrings  $\x^1, \x^2, \ldots, \x^{n/\ell}$, referred to as blocks, each being of size~$\ell$. Suppose that $\x$ is affected by deletions resulting in a string $\y$. We are interested in constructing codes that can detect the exact number of deletions that have occurred in each of the blocks $\x^1, \x^2, \ldots, \x^{n/\ell}$, based on the observed string $\y$. Such codes can equivalently determine the boundaries of each block in~$\y$. Furthermore, these codes also enable localizing deletions and identifying their presence in certain parts of an information string. Thus, they have several potential applications such as coding for segments edit channels~\cite{9691355,A18,L10}, marker-based constructions for coded trace reconstruction~\cite{J20}, file synchronization~\cite{R15}; in addition to detecting the presence of mutations in parts of a genomic sequence and retaining synchronization in sequential transmission. 

Most of the literature has focused on the correction of deletions under the assumption that the codeword boundaries are known at the decoder. Levenshtein~\cite{L66}  derived fundamental limits which show that the optimal number of redundant bits needed to correct $\delta$ worst-case deletions in a binary string of length $n$ is $\Theta(\del\log(n/\del))$. 
Levenshtein also showed that the code constructed by Varshamov and Tenengolts (VT codes)~\cite{VT65} is capable of correcting a single deletion and has an asymptotically optimal redundancy in~$n$. There have been lots of works in the past few years on the classical problem of constructing codes that correct deletions, e.g.,~\cite{B16,GCISIT17,GC,H19,Chen18,SimaIT,SimaSYS}. The state-of-the-art results in \cite{Chen18,SimaIT,SimaSYS} give codes with $\mathcal{O}(\delta \log n)$ redundancy. Some of the recent works also studied the case where the deletions occur in bursts or are localized within certain parts of the codeword, e.g.,~\cite{Sch17,GCAllerton17,GClocalized,A20,B21b}. 

As previously mentioned, the aforementioned works in the literature have a common requirement: the codeword boundaries must be known at the decoder in order to successfully correct the errors. In fact, one can easily show that if multiple codewords of the single deletion-correcting VT code are concatenated and transmitted over a channel that deletes at most one bit in each codeword, then the decoder cannot determine the boundaries of these codewords with certainty. In other words, the decoder cannot detect the number of deletions that have occurred in each codeword. Therefore, the problem that we study in this paper cannot be solved by concatenating codewords that belong to classical deletion-correcting codes.

Codes for detecting deletions have been previously studied in~\cite{K03,P13} under a different definition than the one we use in this paper. The definition used in~\cite{P13} for a deletion detecting code is as follows. A code $\cC\subseteq \mathbb{F}_2^n$ is said to be a $\del$-deletion detecting code in~\cite{P13}, if for any $\x\in \cC$, the process of deleting any $\del'\leq \del$ bits from $\x$ and then appending arbitrary $\del'$ bits at the end, does not generate a codeword in~$\cC$. The authors in~\cite{K03} consider a similar definition and focus on non-binary codes. The main difference between our definition and the definitions in~\cite{K03,P13} is that we require the decoder to detect the exact number of deletions as opposed to only identifying whether deletions have occurred in a block or not. More specifically, our (informal) definition is the following. We say that a code $\cC$ detects up to $\del$ deletions in each of the blocks of \mbox{$\x=\langle \x^1, \x^2, \ldots, \x^{n/\ell} \rangle \in \cC$}, if and only if there exists a decoder that can determine the exact number of deletions that have occurred in each block after $\x$ is affected by worst-case deletions. The difference between the two previous definitions is crucial in the setting of concatenated strings which we consider in this paper. Namely, in the presence of multiple concatenated blocks, detecting the exact number of deletions in a given block allows the decoder to determine the boundary of that block.

Our main contributions and the organization of this paper are summarized as follows. A formal definition of the problem is given in Section~\ref{sec:problem}. In Section~\ref{sec:bounds}, we study the fundamental limits of the problem by deriving lower bounds on the redundancy of deletion-detecting codes. The bounds show that for a codeword $\x=\langle \x^1, \x^2, \ldots, \x^{n/\ell}\rangle\in \mathbb{F}_2^n$, the redundancy required to detect up to $\del$ deletions in each block of size $\ell$ is at least  $2\del(n/\ell -1)$ bits. Furthermore, for a specific family of efficiently decodable codes, called block-by-block decodable codes, the required redundancy is at least $(2\del+1)(n/\ell -1)$ bits. In Section~\ref{sec:detecting_del}, we introduce an explicit marker-based code construction for detecting up to $\del$ deletions per block. The code is encodable and block-by-block decodable in linear time $\cO(n)$, and its redundancy is \mbox{$(2\del+1)(n/\ell-1)$} bits. Thus, the presented construction is optimal among all block-by-block decodable codes, and asymptotically optimal in $\delta$ among all deletion-detecting codes. In Section~\ref{sec:trace}, we use this code to design a novel coded trace reconstruction scheme that builds on the properties of deletion-detecting codes and existing reconstruction algorithms to allow efficient trace reconstruction from a constant number of traces. We conclude with some open problems in Section~\ref{sec:conc}.

\section{Problem Statement and Notation}\label{sec:problem}
We start by introducing some of the notations used throughout the paper.  Let $[n]$ be the set of integers from $1$ to $n$~(inclusive), and let $[i,j]$ be the set of integers from $i$ to $j$ (inclusive). Let $\mathbf{1}^i$ and $\mathbf{0}^j$ denote strings of $i$ consecutive ones and $j$ consecutive zeros, respectively. For a string \mbox{$\mathbf{x}=(x_1,x_2,\ldots,x_n)\in \mathbb{F}_2^n$}, we use $x_i$, \mbox{$i=1,2,\ldots,n$}, to refer to the $i\th$ bit of $\mathbf{x}$. We write $\mathbf{x}_{[i,j]}=(x_i,x_{i+1},\ldots,x_j)$ as the substring of $\mathbf{x}$ which contains the consecutive bits ranging from index $i$ to index $j$. We use the notation $i\mid j$  to indicate that $j$ is divisible by $i$, and $i\nmid j$ to indicate that $j$ is not divisible by $i$. All logarithms in this paper are of base $2$. Following standard notation, $\mathbb{Z}^+=\{1,2,\ldots\}$ is the set of positive integers, $\mathbb{Z}_i=\{0,1,\ldots,i-1\}$ is the set of non-negative integers strictly less than $i\in \mathbb{Z}^+$, and $\mathbb{F}_2^*=\{0,1\}^*$ is the set of finite binary strings.

\begin{definition}
\label{def1}
For a given $\ell \in \mathbb{Z}^+$, such that $\ell \mid n$ and $\ell<n$, the $j\th$ block of $\mathbf{x}\in \mathbb{F}_2^n$ is defined as the substring $$\mathbf{x}^j\triangleq \mathbf{x}_{[1+(j-1)\ell, j\ell]}=(x_{1+(j-1)\ell},x_{2+(j-1)\ell},\ldots,x_{j\ell}),$$
where $j \in [n/\ell]$.
\end{definition}

We use $\langle \mathbf{a},\mathbf{b}\rangle$ to refer to the concatenation of two strings $\mathbf{a}$ and~$\mathbf{b}$. The formal definition of codes that detect deletions in concatenated strings is given in~Definition~\ref{def2}. In this work, we focus on the case where these strings are binary.

\begin{definition}
\label{def2}
Let $\mathcal{C}_{\del}(\ell,n)\subseteq \mathbb{F}_2^n$, such that $\del< \ell \leq n/2$ and $\ell\mid n$, be a code of length $n$ that contains codewords of the form $$\mathbf{x}=\langle \mathbf{x}^1,\mathbf{x}^2,\ldots,\mathbf{x}^{n/\ell}\rangle,$$ where $\mathbf{x}^j \in \mathbb{F}_2^{\ell}$ for all $j\in [n/\ell]$. Suppose that $\mathbf{x}\in \mathcal{C}_{\del}(\ell,n)$ is affected by at most $\del$ deletions in each block, resulting in a string \mbox{$\mathbf{y}\in \mathbb{F}_2^*$}. The code $\mathcal{C}_{\del}(\ell,n)$ is said to detect up to $\del$ deletions per block, if and only if there exists a decoding function $$\mathrm{Dec}(\mathbf{y}) : \mathbb{F}_2^* \to \mathbb{Z}_{\del+1}^{n/\ell},$$ that outputs the exact numbers of deletions that have occurred in each block $\mathbf{x}^j$, for all $\mathbf{x}$ and $\mathbf{y}$.
\end{definition}

Note that the decoding requirement in Definition~\ref{def2} of detecting the {\em exact} number of deletions per block, is equivalent to determining the boundaries of each block in the observed string~$\mathbf{y}$. 
\begin{example}
Let $\x$ be a binary string of length $20$ given by $$\mathbf{x}=\overbrace{01\underline{0}11}^{\mathbf{x}^1}\overbrace{01 001}^{\mathbf{x}^2}\overbrace{0111\underline{1}}^{\mathbf{x}^3}\overbrace{\underline{1}1 101}^{\mathbf{x}^4},$$ where $\x$ is obtained by concatenating the four blocks $\mathbf{x}^1, \mathbf{x}^2, \mathbf{x}^3,$ and $\mathbf{x}^4$ of length $5$ each. Suppose that the bits underlined in $\mathbf{x}$ are deleted, resulting in $$\mathbf{y}=01110100101111101.$$ If $\mathbf{x}\in \cC_{1}(5,20)$, where $\cC_{1}(5,20)$ is a code that follows Definition~\ref{def2} with $\del=1, \ell=5,$ and $n=20$, then the output of the decoder should be $$\mathrm{Dec}(\mathbf{y})=(1,0,1,1).$$ It follows from the output of the decoder that the boundaries of the blocks are as follows $$\mathbf{y}=\overbrace{0111}^{\text{block 1}}|\overbrace{01001}^{\text{block 2}}|\overbrace{0111}^{\text{block 3}}|\overbrace{1101}^{\text{block 4}}.$$
\end{example}
Next, we define a special family of deletion-detecting codes, called block-by-block decodable codes (Definition~\ref{def3}). The main property of this family of codes is that they can be decoded in an efficient sequential manner by processing $\ell$ bits at a time, i.e., without any lookahead.

\begin{definition}
\label{def3}
Let $\cC_{\del}(\ell,n)$ be a code that follows Definition~\ref{def2}. Consider a codeword $\x \in \cC_{\del}(\ell,n)$ that is affected by at most $\del$ deletions in each of its blocks, resulting in $\y$. Let $\alpha_j$ be the starting position of block $j$ in $\y$, with $\alpha_1=1$. Assume that $\alpha_j$ is known, then the code $\cC_{\del}(\ell,n)$ is said to be block-by-block decodable, if and only if there exists a decoder that can output the exact numbers of deletions in any block $\x^j$ by only processing the bits in~$\y_{[\alpha_j,\alpha_j+\ell-1]}$.
\end{definition}

Following standard notation, $f(n) = o(g(n))$ means that $\lim_{n\to \infty} f(n)/g(n) = 0$; $f(n) = \omega(g(n))$ means that $\lim_{n\to \infty} f(n)/g(n) =\infty$; $f(n) = \Omega(g(n))$ means that $f$ is asymptotically bounded below by $c_1g(n)$ for some constant $c_1> 0$; $f(n) = \cO(g(n))$ means that $f$ is asymptotically bounded above by $c_2g(n)$ for some constant $c_2> 0$; \mbox{$f(n)= \Theta(g(n))$} means that $f(n)/g(n)$ asymptotically lies in an interval $[c_1, c_2]$ for some constants $c_1, c_2 > 0$, and $f(n)=\text{poly}(g(n))$ means that $f(n)=g(n)^{\cO(1)}$.

\section{Lower Bounds on the Redundancy of Deletion-detecting Codes}\label{sec:bounds}
In this section, we derive converse results for the problem. Our results are stated in Section~\ref{sec:bresults}, and the proofs are given in Section~\ref{proofL}

\subsection{Results}\label{sec:bresults}
Let $|\mathcal{C}_{\delta}(\ell,n)|$ be the size of the code $\mathcal{C}_{\delta}(\ell,n)$ and \mbox{$r_{\delta}(\ell,n)\triangleq n-\log \left\lvert \mathcal{C}_{\delta}(\ell,n) \right \rvert$} be its redundancy. In Theorem~\ref{thm2}, we give lower bounds on the redundancy of any code $\cC_{\del}(\ell,n)$ that detects up to $\del$ deletions per block~(Definition~\ref{def2}). In Theorem~\ref{thm3}, we specifically consider block-by-block decdodable codes (Definition~\ref{def3}), and give a lower bound on the redundancy of such codes. 

\begin{theorem}
\label{thm2}
For $\del,\ell,n\in \mathbb{Z}^+$, such that $2\del< \ell \leq n/2$ and $\ell \mid n$,  the redundancy $r_\del(\ell,n)$ of any code $\cC_{\del}(\ell,n)\subseteq \mathbb{F}_2^n$ that detects up to $\del$ deletions per block satisfies: \begin{enumerate}[label={\textit{(\roman*)}}] 
\item If $n/\ell=2$, then  $$r_\del(n/2,n) \geq 2\del;$$
\item If $n/\ell \geq 3$ and $2\del \nmid \ell$, then $$r_\del(\ell,n) \geq 2\del(n/\ell -1);$$
\item If $n/\ell \geq 3$ and $2\del \mid \ell$, then $$r_\del(\ell,n) \geq (2\del+\varepsilon)(n/\ell -1) - \varepsilon,$$
where $0<\varepsilon\leq 0.415$ is a function of $\del$ and $\ell$ given by $$\varepsilon_{\del}(\ell)=\log\left(\frac{2^{\ell-2\del}}{2^{\ell-2\del}-1}\right).$$
\end{enumerate}
\end{theorem}


\begin{theorem}
\label{thm3}
For $\del,\ell,n\in \mathbb{Z}^+$,  such that $2\del< \ell \leq n/2$ and $\ell \mid n$,  the redundancy $r_\del(\ell,n)$ of a block-by-block decodable code \mbox{$\cC_{\del}(\ell,n)\subseteq \mathbb{F}_2^n$} that detects up to $\del$ deletions per block satisfies $$r_\del(\ell,n) \geq (2\del+1)(n/\ell -1).$$
\end{theorem}

{\em Discussion:} The result in Theorem~\ref{thm2} shows that in case we have only two concatenated strings, i.e., $n/\ell=2$, then the redundancy of the code must be at least $2\delta$ bits in order to detect up to $\del$ deletions per block. More specifically, it follows from the proof of Theorem~\ref{thm2} given in the next section, that the redundancy must be at least $\del$ in both the first and second blocks, resulting in a total redundancy of at least $2\del$.

In the case where we have three or more blocks and $2\del$ does not divide $\ell$, the redundancy of the code must be at least $2\del(n/\ell -1)$ bits. It also follows from the proof that this lower bound on the code redundancy is an aggregate of the following block-level requirements: \begin{enumerate*}[label={\textit{(\roman*)}}] \item each of the redundancies of the first and last blocks must be at least~$\del$\footnote{Blocks $\x^1$ and $\x^{n/\ell}$ require less redundancy because they have only one ``neighboring'' block with which the boundaries have to be protected.}; \item~the redundancy of every other block must be at least $2\del$ \end{enumerate*}. Whereas in case $2\del$  divides $\ell$, then the redundancy of the code must be strictly greater than $2\del(n/\ell -1)$ bits. 

Furthermore, for the case of block-by-block decodable codes, it follows from the proof of Theorem~\ref{thm3} that the redundancy of each block must be at least $2\del +1$, excluding the first and last block which require at least $\del$ and $\del+1$ redundancy, respectively. 

\subsection{Proofs of Theorem~\ref{thm2} and Theorem~\ref{thm3}} \label{proofL}
The proofs of Theorem~\ref{thm2} and Theorem~\ref{thm3} have a common part where we show that for any codeword $\x \in \cC_\del(\ell,n)$, the last $\del$ bits of the blocks $\x^1,\dots, \x^{n/\ell -1}$ and the first $\del$ bits of blocks $\x^2,\dots,\x^{n/\ell}$ must be predetermined (fixed). This necessary condition gives a preliminary lower bound on the redundancy of any code $\cC_\del(\ell,n)$, with $2\del< \ell \leq n/2$, that is 
\begin{equation}
\label{eq:bound1}
r_\del(\ell,n)\geq 2\del(n/\ell-1).
\end{equation}  
Then, we show that if $n/\ell\geq 3$ and $2\del \mid \ell$, then an additional necessary condition must hold in each block $\x^j$ for \mbox{$j\in [2,n/\ell-1]$}, which gives 
\begin{align}
r_\del(\ell,n)&\geq 2\del(n/\ell -1) + \varepsilon (n/\ell -2), \\
&= (2\del+\varepsilon)(n/\ell -1) - \varepsilon, \label{eq:bound2}
\end{align} 
where $\varepsilon>0$ is a function of $\del$ and $\ell$ to be determined. For Theorem~\ref{thm3}, we specifically consider block-by-block decodable codes defined in Definition~\ref{def3}. We show that for such codes the bound in~\eqref{eq:bound1} can be improved to
\begin{equation}
\label{eq:bound3}
r_\del(\ell,n) \geq (2\del+1)(n/\ell -1).
\end{equation}

\begin{figure*}
\begin{align}\label{eq:combi_1}
\y_1 &= (\cdots,{x_{\ell-\del}^j},{\color{red}x_{\del+1}^{j+1}},\ {\color{red}\cdots},\ {\color{red}x_{3\del}^{j+1}} \ ,x_{3\del+1}^{j+1},\cdots, x_{5\del}^{j+1}, \cdots \cdots \cdots, {\color{violet}x_{\ell-3\del+1}^{j+1}},{\color{violet}\cdots }, {\color{violet}x_{\ell-\del}^{j+1}},\ {\color{blue} x_{\ell-\del+1}^{j+1}},{\color{blue}\cdots}, {\color{blue} x_\del^{j+2}}, x_{\del+1}^{j+2}, \cdots ). \\ 
\y_2 &= (\cdots,{x_{\ell-\del}^j},{\color{blue}x_{\ell-\del+1}^{j}},{\color{blue}\cdots},{\color{blue}x_{\del}^{j+1}},{\color{red}x_{\del+1}^{j+1}},{\color{red}\cdots},{\color{red}x_{3\del}^{j+1}},\cdots \cdots \cdots, {\color{black}x_{\ell-5\del+1}^{j+1}},\cdots, {\color{black}x_{\ell-3\del}^{j+1}}, {\color{violet}x_{\ell-3\del+1}^{j+1}},{\color{violet}\cdots }, {\color{violet}x_{\ell-\del}^{j+1}},x_{\del+1}^{j+2}, \cdots). \label{eq:combi_2}
\end{align}
\begin{equation} \label{eqqf}
    (\underbrace{{\color{red}x_{\del+1}^{j+1}},{\color{red}\cdots},\ {\color{red}x_{3\del}^{j+1}}}_{2\del\text{ bits}}\ , \underbrace{x_{3\del+1}^{j+1},\cdots, x_{5\del}^{j+1}}_{2\del\text{ bits}}, \cdots \cdots \cdots, \underbrace{{\color{violet}x_{\ell-3\del+1}^{j+1}},{\color{violet}\cdots }, {\color{violet}x_{\ell-\del}^{j+1}}}_{2\del\text{ bits}}) ~\neq ~ \langle \  \underbrace{{\color{blue}\mathbf{1}^{\del}},{\color{blue}\mathbf{0}^{\del}}}_{2\del\text{ bits}} \ ,\underbrace{{\color{blue}\mathbf{1}^{\del}},{\color{blue}\mathbf{0}^{\del}}}_{2\del\text{ bits}} \ ,\cdots \cdots \cdots, \underbrace{{\color{blue}\mathbf{1}^{\del}},{\color{blue}\mathbf{0}^{\del}}}_{2\del\text{ bits}} \ \rangle.
\end{equation}
\end{figure*}

We start by showing that for every codeword $\x\in \cC_\del(\ell,n)$, each bit in the last $\delta$ bits $x_{\ell-\delta+1}^j,\ldots,x_\ell^j$ of any block $\x^j$, \mbox{$j\in [n/\ell -1]$}, must be different from all $\del$ bits $x_{1}^{j+1},\dots,x_{\del}^{j+1}$ of block $\x^{j+1}$. Suppose that for some $i_1 \in [\ell-\del+1,\ell]$ and $i_2 \in [\del]$, we have $x_{i_1}^j=x_{i_2}^{j+1}$. Consider the deletion combination where the last $\ell-i_1+1$ bits $x_{i_1}^j,\ldots,x_\ell^j$ are deleted in $\mathbf{x}^j$, and the first $i_2-1$ bits $x_1^{j+1},\ldots,x_{i_2-1}^{j+1}$ are deleted in $\x^{j+1}$ (if $i_2=1$ no bits are deleted in $\x^{j+1}$). The resulting string is of the form
\begin{equation*}
\y_1 = (\cdots,x_{i_1-1}^j,{\color{red}x_{i_2}^{j+1}},x_{i_2+1}^{j+1},\cdots, x_{\del}^{j+1},\cdots, x_\ell^{j+1}, \cdots),
\end{equation*}
where all other blocks are not affected by deletions. Now consider a different deletion combination where the last $\ell-i_1$ bits $x_{i_1+1}^j,\ldots,x_\ell^j$ are deleted in $\mathbf{x}^j$, and the first $i_2$ bits $x_1^{j+1},\ldots,x_{i_2}^{j+1}$ are deleted in $\x^{j+1}$ (if $i_1=\ell$ no bits are deleted in $\x^{j}$). The resulting string is of the form
\begin{equation*}
\y_2 = (\cdots,x_{i_1-1}^j,{\color{blue}x_{i_1}^{j}},x_{i_2+1}^{j+1},\cdots, x_{\del}^{j+1},\cdots, x_\ell^{j+1}, \cdots),
\end{equation*}
where all other blocks are not affected by deletions. Since $x_{i_1}^{j}=x_{i_2}^{j+1}$ by assumption, then we have $\y_1=\y_2$. %
However, $\y_1$ and $\y_2$ correspond to two different deletion combinations (i.e., two different decoder outputs) given by
\begin{align*}
 \mathrm{Dec}(\y_1) &=(0,\dots,0,\underbrace{\ell-i_1 +1}_{j},\underbrace{i_2-1}_{j+1},0,\dots,0) \in \mathbb{Z}_{\delta+1}^{n/\ell},  \\
 \mathrm{Dec}(\y_2) &= (0,\dots,0,\underbrace{\ell-i_1}_{j},\underbrace{i_2}_{j+1},0,\dots,0) \in \mathbb{Z}_{\delta+1}^{n/\ell}.
\end{align*}

Therefore, if for any $i_1 \in [\ell-\del+1,\ell]$ and $i_2 \in [\del]$ we have $x_{i_1}^{j}=x_{i_2}^{j+1}$, then there exist two different decoder outputs that correspond to the same decoder input, i.e., $\y_1=\y_2$ with $\mathrm{Dec}(\y_1)\neq \mathrm{Dec}(\y_2)$. This contradicts the definition of the decoding function given in Definition~\ref{def2}. We conclude that the following conditions are necessary: 
\begin{equation}\label{eq:condition_2}
    x_{i_1}^j\neq x_{i_2}^{j+1},\ \forall \ i_1\in [\ell-\del+1,\ell], i_2\in[\del], \text{and } j\in [n/\ell -1].
\end{equation}

Since we focus on binary codes, the previous constraints imply that the last $\del$ bits of every block $\x^j$, \mbox{$j\in [n/\ell-1]$}, must be equal and must also be different from the first $\del$ bits of the block $\x^{j+1}$, which also should be equal. Namely, the code $\cC_{\del}(\ell,n)$ can have codewords that either satisfy \begin{enumerate*}[label={\textit{(\roman*)}}] \item $\x^j_{[\ell-\del+1,\ell]}=\mathbf{1}^{\del}$ and $\x^{j+1}_{[1,\del]}=\mathbf{0}^{\del}$; or \item $\x^j_{[\ell-\del+1,\ell]}=\mathbf{0}^{\del}$ and $\x^{j+1}_{[1,\del]}=\mathbf{1}^{\del}$. \end{enumerate*}

Next, we show that although the previous statement is true, the code $\cC_{\del}(\ell,n)$ cannot have a pair of codewords \mbox{$\x_1,\x_2\in \cC_{\del}(\ell,n)$}, where $\x_1$ satisfies \textit{(i)} and $\x_2$ satisfies \textit{(ii)}. To prove this, we suppose that there exists such a pair $\x_1,\x_2\in \cC_{\del}(\ell,n)$, and then show that in this case we have two different decoder outputs that correspond to the same input string $\y$, which contradicts the definition of the decoding function in Definition~\ref{def2}.

Consider the following two deletion combinations for a given \mbox{$j\in [n/\ell-1]$}. In the first one, $\del-1$ out of the last $\del$ bits are deleted in $\x_1^j$, the first $\del$ bits are deleted in $\x_1^{j+1}$, and no bits are deleted in other blocks. In the second one, the last $\del$ bits are deleted in $\x_2^{j}$, $\del-1$ out of the first $\del$ bits are deleted in $\x_2^{j+1}$, and no bits are deleted in other blocks. In both combinations, the resulting string is
\begin{equation*}
\y_1 = \y_2 = (\cdots,{x_{\ell-\del}^j},{\color{red}1},x_{\del+1}^{j+1},x_{\del+2}^{j+1},\cdots, x_\ell^{j+1}. \cdots),
\end{equation*}
Since $\y_1$ and $\y_2$ correspond to two different decoder outputs
\begin{align*}
 \mathrm{Dec}(\y_1) &=(0,\dots,0,\underbrace{\del-1}_{j},\underbrace{\del}_{j+1},0,\dots,0) \in \mathbb{Z}_{\delta+1}^{n/\ell},  \\
 \mathrm{Dec}(\y_2) &= (0,\dots,0,\underbrace{\del}_{j},\underbrace{\del-1}_{j+1},0,\dots,0) \in \mathbb{Z}_{\delta+1}^{n/\ell},
\end{align*}
then we have $\y_1=\y_2$ with $\mathrm{Dec}(\y_1)\neq \mathrm{Dec}(\y_2)$ which contradicts Definition~\ref{def2}. 

Hence, we conclude that the first $\del$ bits and the last $\del$ bits in every block $\x^j$, $j\in [2,n/\ell-1]$, must be predetermined, and the last $\del$ bits of $\x^1$ and the first $\del$ bits of $\x^{n/\ell}$ must be predetermined. Therefore, $$|\cC_\del(\ell,n)| \leq 2^{n - 2\del(n/\ell -1)},$$ which gives the lower bound on the redundancy in~\eqref{eq:bound1}.

Given the aforementioned constraints, we show next that for $n/\ell \geq 3$, the bound in~\eqref{eq:bound1} is not achievable if $2\del \mid \ell$, i.e., a redundancy of exactly $2\del(n/\ell -1)$ bits is not sufficient for decoding. 

For $n/\ell \geq 3$ and a given $j\in[n/\ell-2]$, consider the following two deletion combinations. In the first one, the last $\del$ bits are deleted in $\x^j$, the first $\del$ bits are deleted in $\x^{j+1}$, and no other bits are deleted. In the second one, the last $\del$ bits are deleted in  $\x^{j+1}$, the first $\del$ bits are deleted in $\x^{j+2}$, and no other bits are deleted. If $2\del \mid \ell$, the resulting strings $\y_1$ and $\y_2$ are of the forms given in \eqref{eq:combi_1} and \eqref{eq:combi_2}, respectively. As explained previously, since the two deletion combinations are different, it must hold that $\y_1\neq \y_2$. Without loss of generality, assume that $(x_{\ell-\del+1}^{j},\cdots,x_{\del}^{j+1})=\langle \mathbf{1}^{\del},\mathbf{0}^{\del}\rangle$. In this case, it follows from~\eqref{eq:combi_1} and~\eqref{eq:combi_2} that the condition $\y_1\neq \y_2$ is equivalent to~\eqref{eqqf}. By incorporating the condition in~\eqref{eqqf}, the total number of values that each of the $n/\ell-2$ blocks $\x^2,\ldots,\x^{n/\ell-1}$ can take is at most $2^{\ell-2\del}-1$. One can easily show that this implies that the redundancy of each of these blocks must be at least $2\del+\varepsilon_{\delta}(\ell)$, where $$\varepsilon_{\delta}(\ell)\triangleq \log\left(\frac{2^{\ell-2\del}}{2^{\ell-2\del}-1}\right)>0.$$ Note that $\varepsilon_{\delta}(\ell)$ is a decreasing function of $\ell-2\del$, so its value is maximized when $\ell-2\del$ attains its minimum possible value. Since $\ell>2\del$ and $2\del \mid \ell$, we have $\ell-2\del \geq 2\del$. Hence, the minimum possible value of $\ell-2\del$ is equal to $2$, which is attained when $\ell=4\del$ and $\del=1$. Therefore, $\varepsilon_{\delta}(\ell)\leq \varepsilon_{1}(4)=0.415$. Combining the previous derivations with~\eqref{eq:bound2} concludes the proof of Theorem~\ref{thm2}.

Next, we specifically consider block-by-block decodable codes and show that for such codes, in addition to the constraints in~\eqref{eq:condition_2}, the following must hold
\begin{equation}
\label{eq:condition_3}
x_{\del+1}^{j+1} \neq x_{i_1}^j, \forall \ i_1\in [\ell-\del+1,\ell] \text{ and } j\in [n/\ell -1].
\end{equation}
Namely, the constraint in~\eqref{eq:condition_3} extends the necessary conditions in~\eqref{eq:condition_2} to $i_2 \in [\del+1]$. We know from~\eqref{eq:condition_2} that the bits $x_{\ell-\del+1}^j,\ldots,x_{\ell}^j$ have the same values for all $j\in [n/\ell -1]$. Without loss of generality, assume that $\x^j_{[\ell-\delta+1,\ell]}=\mathbf{1}^{\del}$, and suppose that $x_{\del+1}^{j+1}=1$ has the same value as these bits. Consider the following two deletion combinations. In the first one, only one out of the last $\del$ bits is deleted in $\x^j$, and the first $\del$ bits are deleted in $\x^{j+1}$. In the second one, no bits are deleted in $\x^j$. Let $\alpha_j$ be the starting position of block $j$ in $\y$, and assume that the value of $\alpha_j$ is known at the decoder. For both combinations, we have $$\y_{[\alpha_j,\alpha_j+\ell-1]}=\langle \x^j_{[\ell-\del]},\mathbf{1}^{\del}\rangle.$$ Hence, a decoder cannot determine the exact number of deletions in $\x^j$ (0 or 1) by only processing the $\ell$ bits in $\y_{[\alpha_j,\alpha_j+\ell-1]}$. This contradicts the definition of block-by-block decodable codes given in Definition~\ref{def3}. Therefore, the condition in~\eqref{eq:condition_3} is necessary for all block-by-block decodable codes. This condition introduces an additional redundancy of $1$ bit in each of the $n/\ell-1$ blocks $\x^2,\ldots,\x^{n/\ell}$. Thus, to conclude the proof of Theorem~\ref{thm3}, we add $n/\ell -1$ to the RHS in~\eqref{eq:bound1} and obtain the bound in~\eqref{eq:bound3}.

\section{Optimal Deletion-Detecting Code}\label{sec:detecting_del}
In this section, we show that the lower bound in Theorem~\ref{thm3} can be achieved with equality by a marker-based code construction which we present in Section~\ref{del:cons} (Construction~\ref{cons1}). The properties of this optimal code are discussed in Section~\ref{sec:rc}.

\subsection{Results}\label{sec:rc}
In Theorem~\ref{thm1}, we state our result on the explicit code $\cD_{\del}(\ell,n)$ which we construct in Section~\ref{del:cons}. The proof of the theorem is given in Section~\ref{proof1}.

\begin{theorem}
\label{thm1}
For $\del,\ell,n\in \mathbb{Z}^+$, with $2\del< \ell \leq n/2$, let $$\mathbf{x}=\langle \mathbf{x}^1,\mathbf{x}^2,\ldots,\mathbf{x}^{n/\ell}\rangle \in \cD_{\del}(\ell,n).$$ Suppose that $\mathbf{x}$ is affected by at most $\del$ deletions in each of its blocks $\mathbf{x}^1,\ldots,\mathbf{x}^{n/\ell}$. The code $\cD_{\del}(\ell,n)$ given in Construction~\ref{cons1} detects up to $\del$ deletions per block. This code is encodable and block-by-block decodable in linear time $\cO(n)$, and its redundancy is $(2\del+1)(n/\ell-1)$ bits.
\end{theorem}

{\em Discussion:} Theorem~\ref{thm1} shows that the code $\cD_{\del}(\ell,n)$ which we present in Construction~\ref{cons1} is efficiently encodable and decodable and has redundancy $(2\del+1)(n/\ell-1)$ bits. As we will show in Section~\ref{del:cons}, the total redundancy of $\cD_{\del}(\ell,n)$ is an aggregate of: \begin{enumerate*}[label={\textit{(\roman*)}}] \item $2\del+1$ redundant bits per block $\x^j$, for $j\in[2,n/\ell-1]$; \item~$\del$ redundant bits for $\x^1$; \item $\del+1$ redundant bits for $\x^{n/\ell}$. \end{enumerate*} This means that the redundancy per block only depends on $\del$, and is constant in terms of the size of the block $\ell$ and the size of the codeword $n$. Note that in case we want to also {\em correct} $\del$ deletions per block, then the redundancy per block needs to be $\Omega(\del\log (\ell/\del))$~\cite{L66}, i.e., at least logarithmic in $\ell$. 

Furthermore, $\cD_{\del}(\ell,n)$ is block-by-block decodable, so it follows from Theorem~\ref{thm3} that $\cD_{\del}(\ell,n)$ has {\em optimal} redundancy among all block-by-block decodable codes. Theorem~\ref{thm2} gives lower bounds on the redundancy of any code $\cC_{\del}(\ell,n)$ (not necessarily block-by-block decodable) that detects up to $\del$ deletions per block. Also, by comparing the lower bounds in Theorem~\ref{thm2} to the redundancy of the code in Theorem~\ref{thm1}, it is easy see that for a fixed number of blocks, the presented code $\cD_{\del}(\ell,n)$ has an asymptotically optimal redundancy in $\del$ among all families of deletion-detecting codes $\cC_{\del}(\ell,n)$.

\subsection{Code Construction}
\label{del:cons}
The code that we present for detecting up to $\del$ deletions per block is given by the following construction.
\begin{construction}[Code detecting up to $\del$ deletions]
\label{cons1}
For all $\del,\ell,n\in \mathbb{Z}^+$, with \mbox{$2\del< \ell \leq n/2$}, we define the following%
\begin{align*}
\mathcal{A}_{\del}^0(\ell) &\triangleq \big\{\mathbf{x}\in \mathbb{F}_2^{\ell}~\big|~\mathbf{x}_{[1,\del+1]}=\mathbf{0}^{\del+1}\big\},\\
\mathcal{A}_{\del}^1(\ell) &\triangleq \big\{\mathbf{x}\in \mathbb{F}_2^{\ell} ~\big|~ \mathbf{x}_{[\ell-\del+1,\ell]}=\mathbf{1}^{\del}\big\}.
\end{align*}
The deletion-detecting code $\cD_{\del}(\ell,n)\subseteq \mathbb{F}_2^n$ is given by the following set
\begin{equation*}
\left\{
  \langle \mathbf{x}^1,\ldots,\mathbf{x}^{n/\ell}\rangle \;\middle|\;
  \begin{aligned}
  & \mathbf{x}^1\in \mathcal{A}_{\del}^1(\ell),\\
  & \mathbf{x}^j \in \mathcal{A}_{\del}^1(\ell) \cap \mathcal{A}_{\del}^0(\ell), \forall j \in [2, \frac{n}{\ell}-1], \\
  & \mathbf{x}^{n/\ell} \in \mathcal{A}_{\del}^0(\ell).
  \end{aligned}
\right\}.
\end{equation*}
\end{construction}
Before we prove Theorem~\ref{thm1}, we will provide the steps of the decoding algorithm in Section~\ref{del:dec} and give an example for $\del=1$ in Section~\ref{sec:ex}. The encoding algorithm is omitted since encoding can be simply done by setting the information bits to the positions in $[n]$ that are not restricted by Construction~\ref{cons1}. 

%

\subsection{Decoding} 
\label{del:dec}
Suppose that $\mathbf{x}\in \cD_{\del}(\ell,n)$ (with $2\del< \ell \leq n/2$) is affected by at most $\del$ deletions in each block $\mathbf{x}^j$, for all $j\in[n/\ell]$, resulting in $\mathbf{y}\in \mathbb{F}_2^*$. The input of the decoder is $\mathbf{y}$, and the output is \mbox{$(\delta_1,\delta_2,\ldots,\delta_{n/\ell})\in \mathbb{Z}_{\delta+1}^{n/\ell}$}, where $\delta_j \leq \del$ denotes the number of deletions that have occurred in block $j$. 

The decoding is performed on a block-by-block basis, so before decoding block $j$, we know that the previous $j-1$ blocks have been decoded correctly. Therefore, after decoding the first $j-1$ blocks, the decoder knows the correct starting position of block $j$ in $\mathbf{y}$. Let the starting position of block $j\in [n/\ell]$ in $\mathbf{y}$ be $\alpha_j$, with $\alpha_1=1$; and $\forall j\in [n/\ell -1]$ let $$\mathbf{s}^j \triangleq \mathbf{y}_{[\alpha_j+\ell-\del,\alpha_j+\ell-1]}.$$
To decode block $j$, the decoder scans the bits in $\mathbf{s}^j$ from left to right searching for the first occurrence of a $0$ (if any). If $\mathbf{s}^j$ has no zeros, i.e., \mbox{$\mathbf{s}^j=\mathbf{1}^{\del}$}, then the decoder declares that no deletions ($\delta_j=0$) have occurred in block $j$, and sets the starting position of block \mbox{$j+1$} to \mbox{$\alpha_{j+1}=\alpha_j+\ell$}. Else, if the first occurrence of a $0$ in $\mathbf{s}^j$ is at position $\beta_j$, with \mbox{$1\leq \beta_j \leq \del$}, then the decoder declares that $\del_j=\del-\beta_j+1$ deletions have occurred in block $j$, and sets the starting position of block $j+1$ to \mbox{$\alpha_{j+1}=\alpha_j+\ell-\del_j$}. The decoder repeats this process for each block until the first $n/\ell-1$ blocks are decoded. Finally, the decoder checks the length of the last block in $\mathbf{y}$ based on its starting position $\alpha_{n/\ell}$, and outputs $\del_{n/\ell}$ accordingly. Note that this decoder satisfies Definition~\ref{def3} since the index of the last bit in $\mathbf{s}^j$, $\alpha_j+\ell-1$, is $\ell$ positions away from the starting position of the block $j$, $\alpha_j$.
\begin{remark} 
A code equivalent to $ \cD_{\del}(\ell,n)$ can be obtained by flipping all the zeros to ones and vice-versa in the positions that are restricted by Construction~\ref{cons1}. The decoding algorithm described above can be also modified accordingly. However, this decoding algorithm cannot be applied for a code that combines codewords from the two aforementioned constructions. 
\end{remark}

\begin{remark}
    In our error model, we assume the presence of at most $\delta$ worst-case deletions in each block. Under this model, we prove the correctness of the decoding scheme presented above in Section~\ref{proof1}. Note that if the number of deletions exceeds $\delta$ for any given block, the decoding will fail for that block and also all other blocks that are subsequently processed by the block-by-block decoder.
\end{remark}

\subsection{Example for $\del=1$}\label{sec:ex}
Next we give an example of our code for $\del=1, \ell=5,$ and $n=20$. Consider a codeword $\mathbf{x}\in \cD_{1}(5,20)$ given by $$\mathbf{x}=\overbrace{10\underline{1}0\sepr}^{\mathbf{x}^1}\overbrace{\sepb\sepb 11\sepr}^{\mathbf{x}^2}\overbrace{\sepb\sepb01\underline{\sepr}}^{\mathbf{x}^3}\overbrace{\underline{\sepb}\sepb 100}^{\mathbf{x}^4}.$$ Suppose that the bits underlined in $\mathbf{x}$ are deleted, resulting in $$\mathbf{y}=10010011100010100.$$ To determine the number of deletions in the first block, the decoder first examines $\mathbf{s}^1=y_5=0$, which implies that $\beta_1=1$, and therefore declares that $\delta_1=1$ deletion has occurred in $\mathbf{x}^1$. The starting position of block $2$ is thus set to $\alpha_2=5$. Then, since $\mathbf{s}^2=y_9=1$, the decoder declares that no deletions ($\delta_2=0$) have occurred in $\mathbf{x}^2$, and sets the starting position of block $3$ to $\alpha_3=10$. Similarly, we have $\mathbf{s}^3=y_{14}=0$, implying that $\delta_3=1$ deletion has occurred in $\mathbf{x}^3$, $\beta_3=1$, and $\alpha_4=14$. Now since $\ell-1=4$ bits are left for the last block, the decoder declares that $\delta_4=1$ deletion has occurred in $\mathbf{x}^4$. 

\subsection{Proof of Theorem~\ref{thm1}}
\label{proof1}
The redundancy of the code $\cD_{\del}(\ell,n)$ follows from the number of bits that are fixed in Construction~\ref{cons1}, which is \mbox{$(2\del+1)(n/\ell-1)$}. Encoding can be simply done by setting the information bits to the positions in $[n]$ that are not restricted by Construction~\ref{cons1}.
Hence, the complexities of the encoding and decoding algorithms are $\cO(n)$ since they involve a single pass over the bits with constant time operations. Next, we prove the correctness of the decoding algorithm.

Consider a codeword $\mathbf{x} \in \cD_{\del}(\ell,n)$ that is affected by at most $\delta$ deletions per block resulting in a string~$\mathbf{y}$.  Since the decoding is done on a block-by-block basis, it is enough to prove the correctness of the algorithm for the case where we have $n/\ell=2$ concatenated blocks. To prove that the code $\cD_{\del}(\ell,n)$ detects up to $\del$ deletions per block, we use induction on $\del\in \mathbb{Z}^+$, with $2\del< \ell$ and $n=2\ell$.

{\em Base case:} we first prove correctness for $\del=1$. The codeword is given by $$\mathbf{x}=\langle \mathbf{x}^1,\mathbf{x}^2 \rangle=(x_1,\ldots,x_{n/2},x_{n/2+1},\ldots,x_n) \in \cD_1(n/2,n).$$ To decode the first block, the decoder observes $\mathbf{s}^1=y_{n/2}$. Now consider two cases: \begin{enumerate*}[label={\textit{(\roman*)}}] \item no deletion has occurred in $\mathbf{x}^1$; and \item one bit was deleted in $\mathbf{x}^1$. \end{enumerate*} In the first case, based on Construction~\ref{cons1}, we always have \mbox{$\mathbf{s}^1=y_{n/2}=1$}. Therefore, it follows from the decoding algorithm described in Section~\ref{del:dec}, that the decoder can correctly declare that no deletions have occurred in the first block. Now consider the second case mentioned above. It follows from the code construction that the values of the first two bits of $\mathbf{x}^2$ are both~$0$. Hence, it is easy to see that for any single deletion in $\mathbf{x}^1$, and any single deletion in $\mathbf{x}^2$, we always have \mbox{$\mathbf{s}^1=y_{n/2}=0$}. Thus, the decoder can always correctly detect a single deletion in $\mathbf{x}^1$. The number of deletions in the second block is consequently determined based on the starting position of the second block in $\mathbf{y}$ and the number of bits in $\mathbf{y}$ that are yet to be decoded. This concludes the proof for $\del=1$.

{\em Inductive step:} we assume that the code $\cD_{\del-1}(n/2,n)$ detects up to $\del-1$ deletions per block, and prove that $\cD_{\del}(n/2,n)$ detects up to $\del$ deletions per block. Based on Construction~\ref{cons1}, it is easy to see that if \mbox{$\x\in \cD_{\del}(n/2,n)$}, then \mbox{$\x \in \cD_{\del-1}(n/2,n)$}, and hence \mbox{$\cD_{\del}(n/2,n) \subset  \cD_{\del-1}(n/2,n)$}. Therefore, it follows from the inductive hypothesis that $\cD_{\del}(n/2,n)$ detects up to $\del-1$ deletions per block. Next, we prove that $\cD_{\del}(n/2,n)$ can also detect exactly $\del$ deletions per block. Suppose that exactly $\del$ deletions occur in~$\mathbf{x}^1$. To decode the first block, the decoder scans the bits of $\mathbf{s}^1=\mathbf{y}_{[\ell-\del+1,\ell]}$ from left to right searching for the first occurrence of a $0$, as explained in Section~\ref{del:dec}. It follows from the code construction that for any $\del$ deletions in $\mathbf{x}^1$, the first bit of $\mathbf{s}^1$ is always $0$, i.e., $y_{\ell-\del+1}=0$. To see this, notice that: \begin{enumerate*}[label={\textit{(\roman*)}}] \item for any $\del$ deletions in $\mathbf{x}^1$, the first bit belonging to the second block in $\mathbf{y}$ will shift $\del$ positions to the left; and \item since $\mathbf{x}_{[n/2+1,n/2+\del+1]}=\mathbf{0}^{\del+1}$, and given that we consider at most $\del$ deletions in $\mathbf{x}^2$, then the first bit belonging to the second block in $\mathbf{y}$ is always a $0$. \end{enumerate*} Hence, for any $\del$ deletions in $\mathbf{x}^1$, and for any $\del$ or fewer deletions in $\mathbf{x}^2$, we have $y_{\ell-\del+1}=0$. It follows from the decoding algorithm that the decoder in this case declares that $\del$ deletions have occurred in~$\mathbf{x}^1$. Furthermore, the number of deletions in the second block is consequently determined based on the starting position of the second block in $\mathbf{y}$ and the number of bits in $\mathbf{y}$ that are yet to be decoded.

\section{Application to Trace Reconstruction}\label{sec:trace}
In this section, we present one possible application of deletion-detecting codes, where we apply the marker-based code in Construction~\ref{cons1} to trace reconstruction.
\subsection{Problem Statement and Related Work}
In trace reconstruction, the goal is to reconstruct an unknown sequence $\mathbf{x}\in \mathbb{F}_2^n$, given random traces of $\mathbf{x}$, where each trace is generated independently as the output of a random channel. We focus on the case of deletion channels, in which each bit of $\mathbf{x}$ is deleted independently and with probability~$p$. The objective is to devise efficient algorithms that allow reconstructing $\mathbf{x}$ from a few traces. There have been lots of works in the literature that study trace reconstruction in the case where $\mathbf{x}$ is uncoded, e.g., \cite{levenshtein1997reconstruction, n1,n2,Batu,holenstein2008trace,mcgregor2014trace,de2017optimal,nazarov2017trace, hartung2018trace,peres2017average,holden2018subpolynomial,chen2020polynomial}.  

The problem of trace reconstruction arises naturally in the data retrieval phase of DNA storage. Namely, state-of-the-art DNA sequencing methods often output several erroneous copies of the stored data. This has motivated a series of recent works to study the problem of trace reconstruction in the coded setting~\cite{J20,B20,AbroshanTrace,srinivasavaradhan2021trellis,maarouf2022concatenated}, where one is allowed to encode the sequence $\mathbf{x}$ before observing multiple traces of it\footnote{Encoding data in DNA storage can be achieved by employing DNA synthesis strategies that enable encoding arbitrary digital information and storing them in DNA~\cite{church2012next,goldman2013towards,yazdi2017portable}.}. 

Data encoding is a natural and well-known strategy to improve the reliability of storage systems. Since we focus on deletion channels, the baseline we consider is classical deletion-correcting codes (e.g.~\cite{B16,GC,H19,Chen18,SimaIT,SimaSYS}), which would only require a single trace to reconstruct the stored data. The optimal redundancy (i.e., storage overhead) to correct $d$ deletions in a sequence of length $n$ using a deletion-correcting code is $\Theta(d \log (n/d))$~\cite{L66}. Furthermore, most of the existing deletion codes in the literature have complex encoding and decoding algorithms. Therefore, the goal in coded trace reconstruction is to exploit the presence of multiple traces which arises naturally in applications like DNA storage, to design codes that:  \begin{enumerate*}[label={\textit{(\roman*)}}] \item have a redundancy that grows strictly slower than the optimal redundancy of deletion-correcting codes; and/or \item can be reconstructed from a few traces using simple and efficient reconstruction algorithms. \end{enumerate*}

The work by Abroshan {\em et al.} in~\cite{AbroshanTrace} studies coded trace reconstruction in the setting where each trace is affected by~$d$ deletions, where $d$ is fixed with respect to the sequence length~$n$, and the positions of the $d$ deletions are uniformly random. The high-level idea of the approach in~\cite{AbroshanTrace} is to concatenate a series of Varshamov-Tenengolts (VT)~\cite{VT65} codewords, which are single deletion-correcting codes that have an asymptotically optimal redundancy in $n$; in addition to efficient linear time encoding and decoding algorithms. The number of concatenated VT codewords is chosen to be a constant that is strictly greater than $d$ so that the average number of deletions in each VT codeword is strictly less than one. An efficient reconstruction algorithm is then used that reconstructs each VT codeword by either finding a deletion-free copy of it among the traces or by decoding a single deletion using the VT decoder. Since the length of each codeword is a constant fraction of~$n$, and since the redundancy of VT codes is logarithmic in the blocklength, then it is easy to see that the resulting redundancy of the code in~\cite{AbroshanTrace} is $\Theta(\log n)$. The works in~\cite{J20} and \cite{B20} study coding for trace reconstruction in the regime where the deletion probability $p=\epsilon$ is fixed, i.e., a linear number of bits are deleted on average. For this regime, the authors in~\cite{J20} introduced binary codes with rate $1-\cO(\frac{1}{\log n})$ that can be efficiently reconstructed from $\log^cn$ traces, with $c>1$. The authors in~\cite{B20} prove the existence of binary codes with rate $1-\cO(\frac{1}{\log n})$ that can be reconstructed from $\exp \left( (\log \log n)^{1/3} \right)$ traces. 

\subsection{Contributions}
We focus on coded trace reconstruction for binary sequences affected by deletions only, as a first step towards designing codes that are robust against the various types of errors that are experienced in DNA storage. We consider the random deletion model in which each bit is deleted independently with probability $p$. We focus on the regime where $p=k/n^{\alpha}$, where $k>1$ and $\alpha \in (0.5,1]$ are constants. We use the deletion-detecting code introduced in Section~\ref{del:cons} (Construction~\ref{cons1}) to design a novel code $\mathcal{C}'\subseteq \mathbb{F}_2^n$ for trace reconstruction. The contributions that we present are the following.
\begin{itemize}[leftmargin=*]
\item For all $\alpha\in (0.5,1]$, the redundancy of our code is $o(n^{1-\alpha} \log n)$, i.e., grows strictly slower than the optimal redundancy of a deletion-correcting code, which is $\Theta(n^{1-\alpha} \log n)$ for $d=\Theta(np)=\Theta(n^{1-\alpha})$. Furthermore, for the particular regime where the average number of deletions is fixed, i.e., $\alpha=1$ and $p=k/n$, we show that our code can efficiently reconstruct any codeword $\mathbf{x}\in \mathcal{C}'$ with any redundancy that is $\omega(1)$. This regime is similar to the regime where the number of deletions per trace is fixed, which was studied in~\cite{AbroshanTrace}. For this regime, our code has a lower redundancy compared to the $\Theta(\log n)$ redundancy of the code introduced in~\cite{AbroshanTrace}.

\item The reconstruction algorithm that we propose runs in linear time $\cO(n)$.

\item We provide a theoretical guarantee on the performance of our code in the form of an upper bound on the probability of reconstruction error. This bound shows that for all $\alpha\in (0.5,1]$, and for any codeword $\mathbf{x}\in \mathcal{C}'$, the probability of reconstruction error vanishes asymptotically in $n$ for a {\em constant} number of traces $t=\cO(1)$. 

\item We evaluate the numerical performance of our code in terms of the edit distance error, which was adopted as a performance metric in several recent related works, e.g.,~\cite{Diggavi, Eitan, Approximate}. We compare the performance of our code to the code in~\cite{AbroshanTrace} and to a simple coded version of the Bitwise Majority Alignment (BMA) reconstruction algorithm~\cite{Batu}. The simulation results show that our code has a low edit distance error for a small number of traces and outperforms both the code in~\cite{AbroshanTrace} and a simple coded version of the BMA algorithm. 
\end{itemize}

\subsection{Notation}
We consider the following deletion model. For a given input sequence $\mathbf{x}=(x_1,x_2,\ldots,x_n)\in \mathbb{F}_2^n$, a deletion probability $p=k/n^{\alpha}$, where $k>1$ and $\alpha\in (0.5,1]$ are constants, and an integer $t$, the channel returns $t$ traces of~$\mathbf{x}$. Each trace of $\mathbf{x}$, denoted by $\mathbf{y}_i \in \mathbb{F}_2^*$, $i=1,\ldots,t$, is obtained by sending $\mathbf{x}$ through a deletion channel with deletion probability $p$, i.e., the deletion channel deletes each bit $x_i$ of $\mathbf{x}$, $i=1,\ldots,n$, independently with probability $p$, and outputs a subsequence of $\mathbf{x}$ containing all bits of $\mathbf{x}$ that were not deleted in order. The $t$ traces $\mathbf{y}_1,\ldots,\mathbf{y}_t$, are independent and identically distributed as outputs of the deletion channel for input~$\mathbf{x}$. The inputs of the reconstruction algorithm are the $t$ traces $\mathbf{y}_1,\ldots,\mathbf{y}_t$, the codeword length $n$, and the channel parameters $k$ and $\alpha$. Let $\hat{\mathbf{x}}\in \mathbb{F}_2^n$ denote the output of the reconstruction algorithm. The probability of reconstruction error is defined by $P_e\triangleq \Pr(\hat{\mathbf{x}} \neq \mathbf{x})$, and the edit distance error is defined by the random variable $L_d \triangleq \lev(\mathbf{x},\hat{\mathbf{x}})$, where $\lev(.,.)$ denotes the Levenshtein distance~\cite{L66} between two sequences, and the randomness in $P_e$ and $L_d$ is over the deletion process. 

\subsection{Preliminaries}
Next, we present a high-level overview of our code construction and explain the main ideas behind the techniques that we use. Our formal results are given in Section~\ref{code:full}.

\subsubsection{Overview of Code Construction} The high-level idea of our code construction is the following. We construct a codeword $\mathbf{x}\in \mathbb{F}_2^n$, by concatenating $n/\ell$ blocks $\mathbf{x}^1,\mathbf{x}^2,\ldots,\mathbf{x}^{n/\ell}$ (Definition~\ref{def1})  each of length $\ell$, where $\ell \in \mathbb{Z}^+$ is a parameter of the code\footnote{For the sake of simplicity, we will assume that $\ell$ divides $n$ and drop this assumption later in Section~\ref{code:full}.}. We construct each block in \mbox{$\mathbf{x}=\langle \mathbf{x}^1,\mathbf{x}^2,\ldots,\mathbf{x}^{n/\ell}\rangle$} such that it satisfies the following two properties. First, the blocks satisfy a {\em run-length-limited} constraint, i.e., the length of each run of $0$'s or $1$'s is limited by a maximum value. Second, each block has a small number of fixed bits at the beginning and at the end of the block that correspond to the markers specified by Construction~\ref{cons1}. 

The goal of introducing the markers is to recover the boundaries of the blocks at the decoder. Determining the block boundaries allows us to obtain the traces corresponding to each block. Consequently, we can subdivide the trace reconstruction problem into smaller subproblems. These subproblems are ``easier" in the sense that the expected number of deletions per block is smaller than the total number of deletions in $\mathbf{x}$, for the same number of traces. This increases the chance of reconstructing a higher number of blocks correctly, thereby resulting in a low edit distance error as we show in Section~\ref{sec:simul}. Moreover, the markers can also prevent potential reconstruction errors from propagating to all parts of the codeword. To understand the intuition behind the subdivision, we present the following example.

Consider a naive trace reconstruction algorithm that attempts to reconstruct a sequence $\mathbf{x}$ by finding a deletion-free copy of it among the traces. In the uncoded setting with a small deletion probability, we expect to observe a small number of deletions per trace. We also expect that many parts (i.e., blocks) of the trace will be deletion-free. Hence, if we use a deletion-detecting code (Definition~\ref{def2}) that enables recovering the boundaries of each block, then we have a higher chance of reconstructing $\mathbf{x}=\langle \mathbf{x}^1,\mathbf{x}^2,\ldots,\mathbf{x}^{n/\ell}\rangle$ by finding one deletion-free copy of each block among the traces. 

For instance, consider the setting where the decoder has \mbox{$t=2$} traces of a binary sequence $\mathbf{x}$ of length $n$. Suppose that the first trace $\mathbf{y}_1$ is affected by a single deletion at position $i_1$, with $i_1<n/2$, and the second trace $\mathbf{y}_2$ is affected by a single deletion at position $i_2$, with $i_2>n/2$. Therefore, the decoder has
\begin{align*}
\mathbf{y}_1 &= (x_1,...,x_{i_1-1},x_{i_1+1},...,x_{n/2},............................,x_n), \\
\mathbf{y}_2 &= (x_1,............................,x_{n/2},...,x_{i_2-1},x_{i_2+1},...,x_n).
\end{align*}
If $\mathbf{x}$ is uncoded, the reconstruction is unsuccessful using the naive algorithm since there is no deletion-free copy of $\mathbf{x}$ in both traces. Now suppose that $\mathbf{x}\in \cC_{1}(n/2,n)$, where $\cC_{1}(n/2,n)$ is a code that follows Definition~\ref{def2}, i.e., can detect up to one deletion in each of the two blocks of size $\ell=n/2$. Then, $\mathbf{x}$ can be successfully reconstructed since knowing the boundaries of each block would allow us to obtain a deletion-free copy of the first block in the second trace, and a deletion-free copy of the second block in the first trace. Note that in general, we may have cases where we cannot find a deletion-free copy of every block. However, we expect that the subdivision will still allow us to find a deletion-free copy of most blocks, resulting in a lower edit distance error compared to the uncoded setting.

Motivated by the previous example, to reconstruct $\mathbf{x}$ at the decoder, we propose the following approach. The decoder obtains $t$ independent traces of $\mathbf{x}$ resulting from $t$ independent deletion channels. To reconstruct $\mathbf{x}$, the decoder first uses the underlying deletion-detecting code to recover the block boundaries of each trace, as previously explained in Section~\ref{sec:detecting_del}. The decoder then uses the Bitwise Majority Alignment (BMA) algorithm, which we explain next, to reconstruct each block from its corresponding traces. To finalize the decoding, the recovered blocks are concatenated in order to reconstruct~$\mathbf{x}$. 

\subsubsection{Bitwise Majority Alignment}
\label{ch4:bma} 
The bitwise majority alignment (BMA) algorithm was first introduced in~\cite{Batu} as an algorithm that reconstructs an {\em uncoded} binary sequence \mbox{$\mathbf{x}\in \mathbb{F}_2^n$} from $t$ traces. In our approach, we first determine the boundaries of each block of size $\ell$ and then apply the BMA reconstruction algorithm at the level of each block. Hence, the input of the algorithm is a $t \times \ell$ binary matrix which consists of the $t$ traces corresponding to a certain block. Since the length of some of the received traces may be smaller than $\ell$ due to the deletions, the traces in the input of the algorithm are padded to length $\ell$ by adding a special character (other than $0$ or $1$). The main idea of the algorithm follows a majority voting approach, where each bit is reconstructed based on its value in the majority of the traces. Namely, for each trace, a pointer is initialized to the leftmost bit, and the value of the bit in the reconstructed sequence is decided based on the votes of the majority of the pointers. Then, the pointers corresponding to the traces that voted with the majority are incremented by one, while other pointers remain at the same bit position. Thus, the reconstruction process scans the bits of the traces from left to right, and the pointers may possibly be pointing to different bit positions at different states of the algorithm. In the end, the algorithm outputs a single reconstructed sequence of length $\ell$ bits. 

Let $q(j)$ be the pointer corresponding to trace $j$, $j=1,\ldots,t$, where $q(j)=i$, $i=1,\ldots,\ell$, means that the pointer corresponding to trace $j$ is pointing to the bit at position $i$. The detailed algorithm is given in Algorithm~\ref{algo_disjdecomp}.

\IncMargin{1em}
\begin{algorithm}[h]
\SetKwInOut{Input}{input}\SetKwInOut{Output}{output}
\Input{A binary matrix $T$ of size $t\times \ell$}
\Output{A binary sequence $y$ of size $\ell$}
\BlankLine
Let $q(j) \leftarrow 1$ for all $j=1,\ldots,t$\;
\For{$i\leftarrow 1$ \KwTo $\ell$}{
	Let $b$ be the majority over all $j$ for $T(j,q(j))$\;
	$y(i)\leftarrow b$\;
	\For{$j\leftarrow 1$ \KwTo $t$}{
		\If{$q(j)==b$}{
			$q(j)\leftarrow q(j)+1$\; 
		}
	} 
}
\caption{Bitwise Majority Alignment}\label{algo_disjdecomp}
\end{algorithm}

The authors in~\cite{Batu} showed that for a deletion probability that satisfies $p=\mathcal{O}(1/\sqrt{\ell})$, the BMA algorithm (Algorithm~\ref{algo_disjdecomp}) can reconstruct an arbitrary (uncoded) binary sequence of size $\ell$, with high probability\footnote{The probability of reconstruction error vanishes asymptotically in $\ell$, where the probability is taken over the randomness of the deletion process.}, from $t=\cO(\ell \log \ell)$ traces. The BMA algorithm has a poor performance for some sequences that have {\em long} runs of $0$'s or $1$'s~\cite{Batu}. For example, consider a sequence that starts with a long run of $0$'s. Due to the deletions, the number of $0$'s observed for this run differs from one trace to another. While scanning from left to right, at some point before the end of this run, the majority of the traces could vote for a~$1$. This will lead to splitting the run resulting in an erroneous reconstruction. The successful reconstruction of such sequences with long runs requires a number of traces that is significantly larger than that of {\em run-length-limited} sequences. In fact, an intermediary result in~\cite{Batu} showed that for {\em run-length-limited} sequences where the length of any run is limited to a maximum value of $\sqrt{\ell}$, the number of traces required by the BMA algorithm is $t=\mathcal{O}(1)$. However, in the context of uncoded worst-case trace reconstruction considered in~\cite{Batu}, the BMA algorithm would still require $t=\cO(\ell \log \ell)$ traces so that the probability of reconstruction error (taken over the randomness of the deletion process) vanishes asymptotically for {\em any} binary sequence. In this section, one of our approaches is to leverage coded trace reconstruction to decrease the required number of traces by restricting our code to codewords of the form $\mathbf{x}=\langle \mathbf{x}^1,\mathbf{x}^2,\ldots,\mathbf{x}^{n/\ell}\rangle\in \mathbb{F}_2^n$, where the length of a run in any block of size $\ell$ is upper bounded by $\sqrt{\ell}$.

\begin{remark}
\label{remBMA}
A straightforward coded version of the BMA algorithm can be obtained by restricting the input codewords $\mathbf{x}\in \mathbb{F}_2^n$  to sequences that do not contain a run that is greater than $\sqrt{n}$. Note that this simple variation of the BMA reconstruction algorithm has not been discussed in the literature, but based on an intermediary result in~\cite{Batu}, one can easily see that this coded version of the BMA algorithm would require a constant number of traces. However, our results in Section~\ref{sec:simul} show that there is a significant difference in terms of the edit distance error between our proposed code and this simple coded version of the BMA algorithm. This highlights the importance of the step of introducing the markers to enable subdividing the problem into smaller subproblems.
\end{remark}

\subsection{Theoretical Results}
\label{code:full}

Let $\mathcal{L}^{n}(f(n))\subseteq \mathbb{F}_2^{n}$ be the set of {\em run-length-limited} binary sequences of length $n$ such that the length of any run of $0$'s or $1$'s in a sequence $\mathbf{x}\in \mathcal{L}^{n}(f(n))$ is at most $f(n)$.

\begin{construction}[Code for Trace Reconstruction]
\label{cons:full}
Consider a given deletion probability \mbox{$p=k/n^{\alpha}<1$}, where $k>1$ and \mbox{$\alpha\in(0.5,1]$} are constants. The code consists of codewords of the form $\mathbf{x}=\langle \mathbf{x}^1,\mathbf{x}^2,\ldots,\mathbf{x}^{\lceil n/\ell \rceil}\rangle$, where the length of each block is set to $\ell=\lfloor 1/p \rfloor$. Let $\delta \in \mathbb{Z}^+$, with $\delta>1$, be a code parameter that represents a strict upper bound on the number of deletions that can be detected in each block $\mathbf{x}^m$, $m\in \{1,\ldots, \lceil n/\ell \rceil\}$, i.e., the code can detect up to $\delta-1$ deletions per block. For $\ell=\lfloor 1/p \rfloor>\delta^2$ we define the code
$$\mathcal{C}'_{\delta}(n)\triangleq \left \{ \mathbf{x}\in \mathbb{F}_2^n~|~\mathbf{x}\in \mathcal{D}_{\delta-1}(\lfloor 1/p \rfloor,n) \cap \mathcal{L}^{n}(\sqrt{\lfloor 1/p \rfloor}) \right \},$$
where $\mathcal{D}$ is the deletion-detecting code defined in Construction~\ref{cons1}.
\end{construction}

The steps of the reconstruction algorithm for a sequence $\mathbf{x}\in\mathcal{C}'_{\delta}(n)$ are as follows.
\begin{enumerate}
\item Recover the block boundaries in all $t$ traces based on the decoding algorithm of  $\mathcal{D}_{\delta-1}(\lfloor 1/p \rfloor,n)$ (Section~\ref{del:dec}).
\item Apply the BMA algorithm (Algorithm~\ref{algo_disjdecomp}) to reconstruct each block.
\item Concatenate the reconstructed blocks to obtain $\mathbf{x}$.
\end{enumerate}

\begin{theorem}
\label{thm:main1}
For any deletion probability \mbox{$p=k/n^{\alpha}$} with $0<p<0.5$, where $k>1$ and \mbox{$\alpha\in(0.5,1]$} are constants, the code $\mathcal{C}'_{\delta}(n)\subseteq \mathbb{F}_2^n$ (Construction~\ref{cons:full}) can be reconstructed from \mbox{$t=\cO(1)$} traces in linear time $\cO(n)$. The redundancy of the code $r_{\mathcal{C}'}(\delta)$ satisfies $$(kn^{1-\alpha}-1)(2\delta-1)+o(1)\leq r_{\mathcal{C}'}(\delta) < 2kn^{1-\alpha}(2\delta-1)+o(1).$$ Let $p(n)=\cO( \text{{\em poly}} (n))$, for $\delta=\lceil \delta^*(p(n)) \rceil=o(\log n)$ with $$\delta^*(p(n))=\frac{2\ln \left( \sqrt{e}n^{1-\alpha}p(n) \right)}{W\left(2e \ln \left( \sqrt{e} n^{1-\alpha}p(n)  \right) \right)},$$ where $W(.)$ is the Lambert $W$ function, the probability of reconstruction error for {\em any} codeword $\mathbf{x}\in \mathcal{C}'_{\delta}(n)$ satisfies $$P_e < \cO\left(n^{1-2\alpha} + \frac{1}{p(n)}\right).$$
 \end{theorem}
 \begin{corollary}
 \label{cor1}
 For any deletion probability \mbox{$p=k/n^{\alpha}$} with \mbox{$0<p<0.5$}, where $k> 1$ and \mbox{$\alpha\in(0.5,1)$} are constants, the code defined in Construction~\ref{cons:full} can be reconstructed from $\cO(1)$ traces in linear time $\cO(n)$, with redundancy $$r_{\mathcal{C}}
=\Theta \left(n^{1-\alpha} \frac{\log n}{\log \log n} \right),$$ and a probability of reconstruction error that satisfies $$P_e< \cO(n^{1-2\alpha}),$$ for any codeword.
 \end{corollary}
 \begin{corollary}
 \label{cor2}
For any deletion probability \mbox{$p=k/n$} with \mbox{$0<p<0.5$}, where $k> 1$ is a constant, the code defined in Construction~\ref{cons:full} can be reconstructed from $\cO(1)$ traces in linear time $\cO(n)$, with a probability of reconstruction error that vanishes asymptotically in $n$ for any redundancy $r_{\mathcal{C}}=\omega(1)$.
 \end{corollary}
 {\em Discussion:} Theorem~\ref{thm:main1} shows that the code defined in Construction~\ref{cons:full} can be efficiently reconstructed from a constant number of traces. Also, the result expresses the redundancy and the probability of reconstruction error in terms of the code parameters and shows a relationship (trade-off) between these two code properties. Corollaries~\ref{cor1} and \ref{cor2} follow from Theorem~\ref{thm:main1}. Corollary~\ref{cor1} shows that for $\alpha\in(0.5,1)$ the redundancy is $$r_{\mathcal{C}}=\Theta \left(n^{1-\alpha}\frac{\log n}{\log \log n}\right)=o(n^{1-\alpha}\log n),$$ and the probability of reconstruction error vanishes asymptotically in $n$ as a polynomial of degree $1-2\alpha$. Note that the aforementioned redundancy grows strictly slower than the optimal redundancy of a deletion-correcting code (with $t=1$ trace). Moreover, the complexity of the reconstruction algorithm is linear in~$n$. Corollary~\ref{cor2} shows that for $\alpha=1$, a vanishing probability of reconstruction error can be achieved with  a constant number of traces and any $\omega(1)$ redundancy.

\subsection{Simulation Results}
\label{sec:simul}
In this section, we evaluate the numerical performance of our code $C_{\delta}'(n)$ defined in Construction~\ref{cons:full} in terms of the edit distance error $L_d=\lev (\hat{\mathbf{x}},\mathbf{x})$ \cite{Diggavi, Eitan, Approximate}. We provide simulation results where we compare the normalized edit distance error $\lev (\hat{\mathbf{x}},\mathbf{x})/n$ of our code to: \begin{enumerate*}[label={\textit{(\roman*)}}] \item the coded version of the BMA algorithm (Remark~\ref{remBMA}) for multiple values of $t, n$ and \mbox{$\alpha\in(0.5,1]$}; and \item the code in~\cite{AbroshanTrace} for $\alpha=1$ and multiple values of $t$. \end{enumerate*}

\begin{figure}[!htb]
\begin{center}
\includegraphics[width=\linewidth]{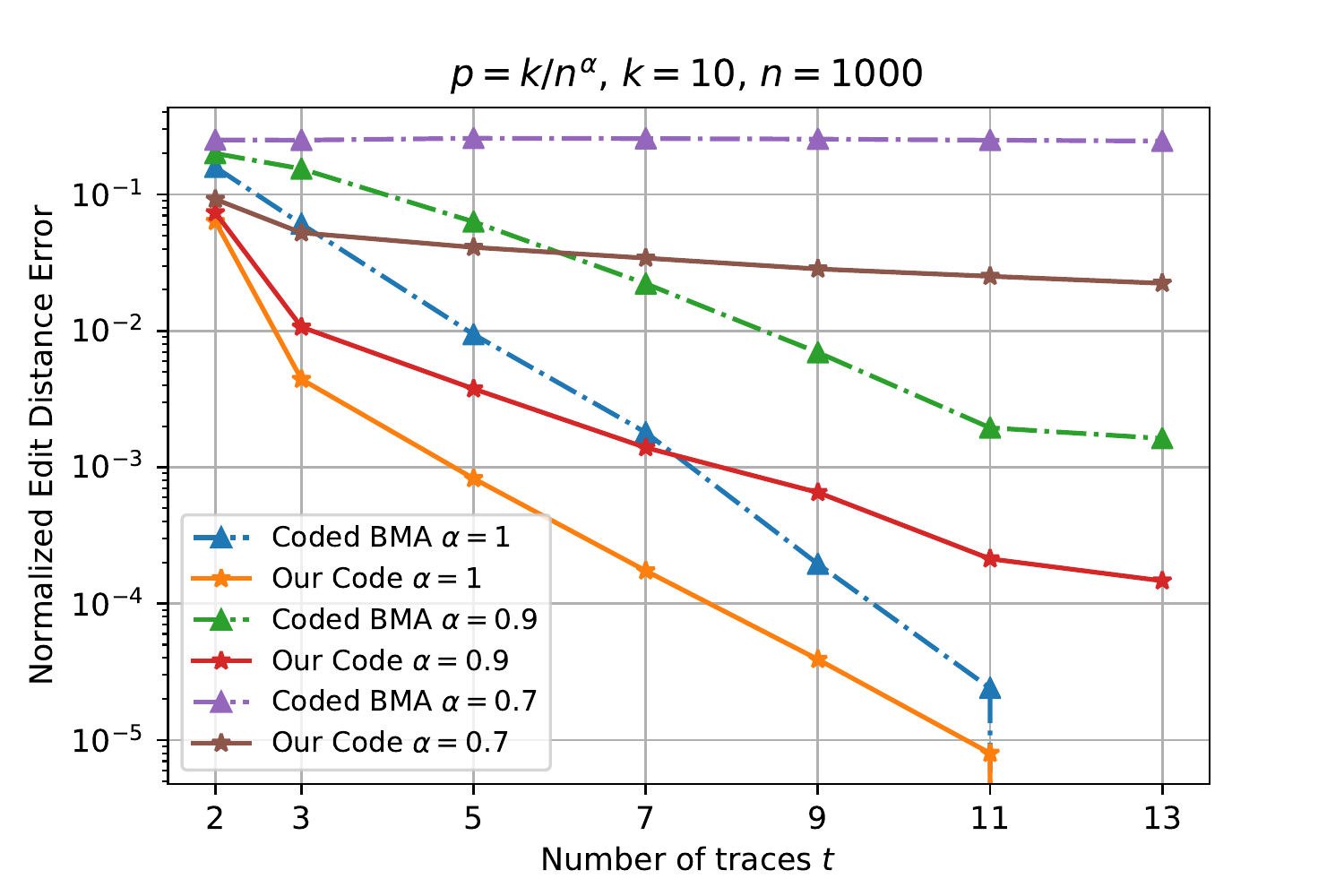}
  \caption{Normalized edit distance error as a function of the number of traces $t$ for codeword length \mbox{$n=1000$} and deletion probability $p=k/n^{\alpha}$, with $k=10$ and $\alpha\in\{0.7, 0.9, 1\}$.}\label{fig:1}
  \end{center}
\end{figure}

\begin{figure}[!htb]
\begin{center}
\includegraphics[width=\linewidth]{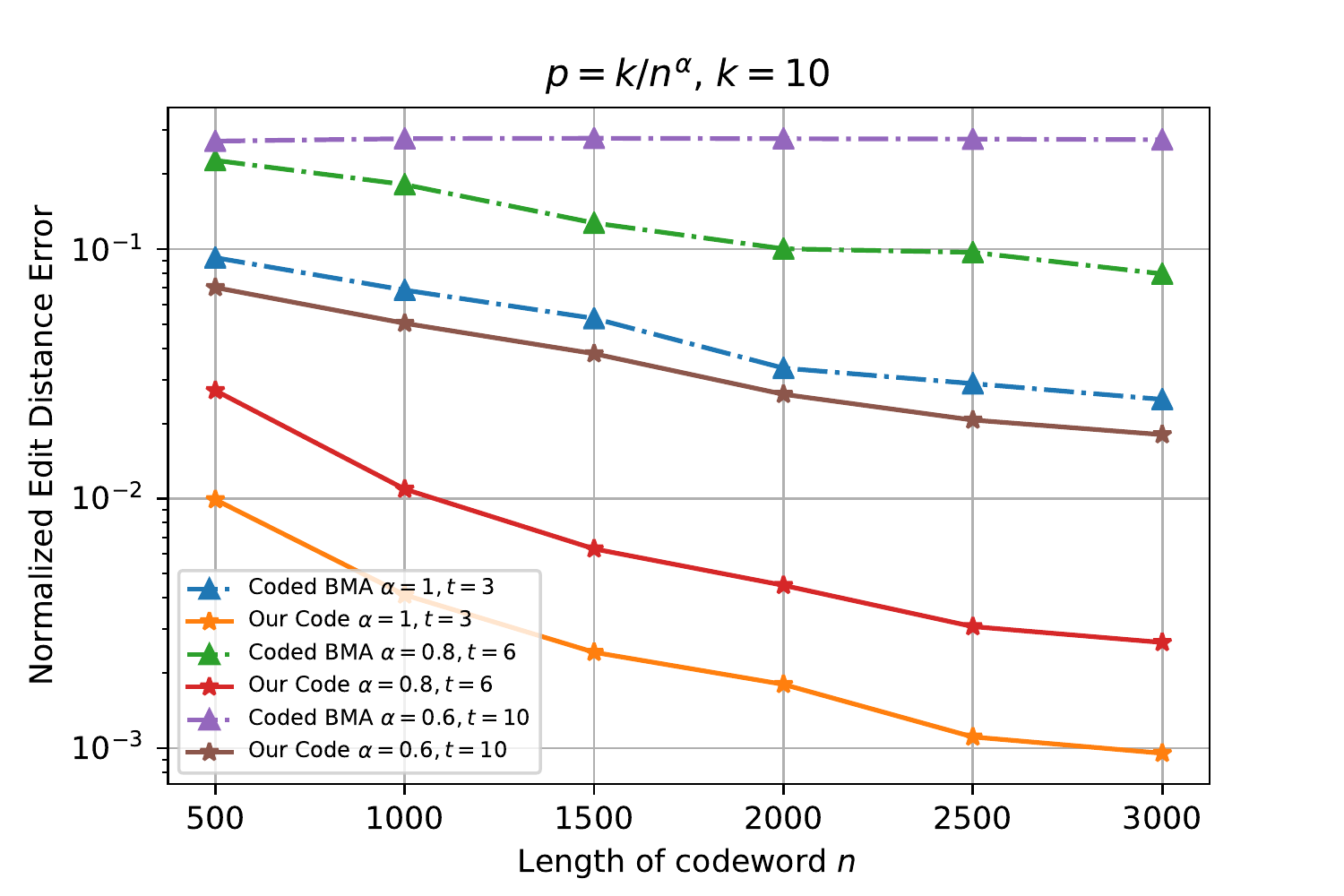}
  \caption{ Normalized edit distance error as a function of the codeword length $n$, for deletion probability $p=k/n^{\alpha}$, with $k=10$ and $(\alpha,t)\in \{(1,3), (0.8, 6), (0.6,10)\}$.}\label{fig:2}
  \end{center}
\end{figure}

\begin{figure}[!htb]
\begin{center}
\includegraphics[width=\linewidth]{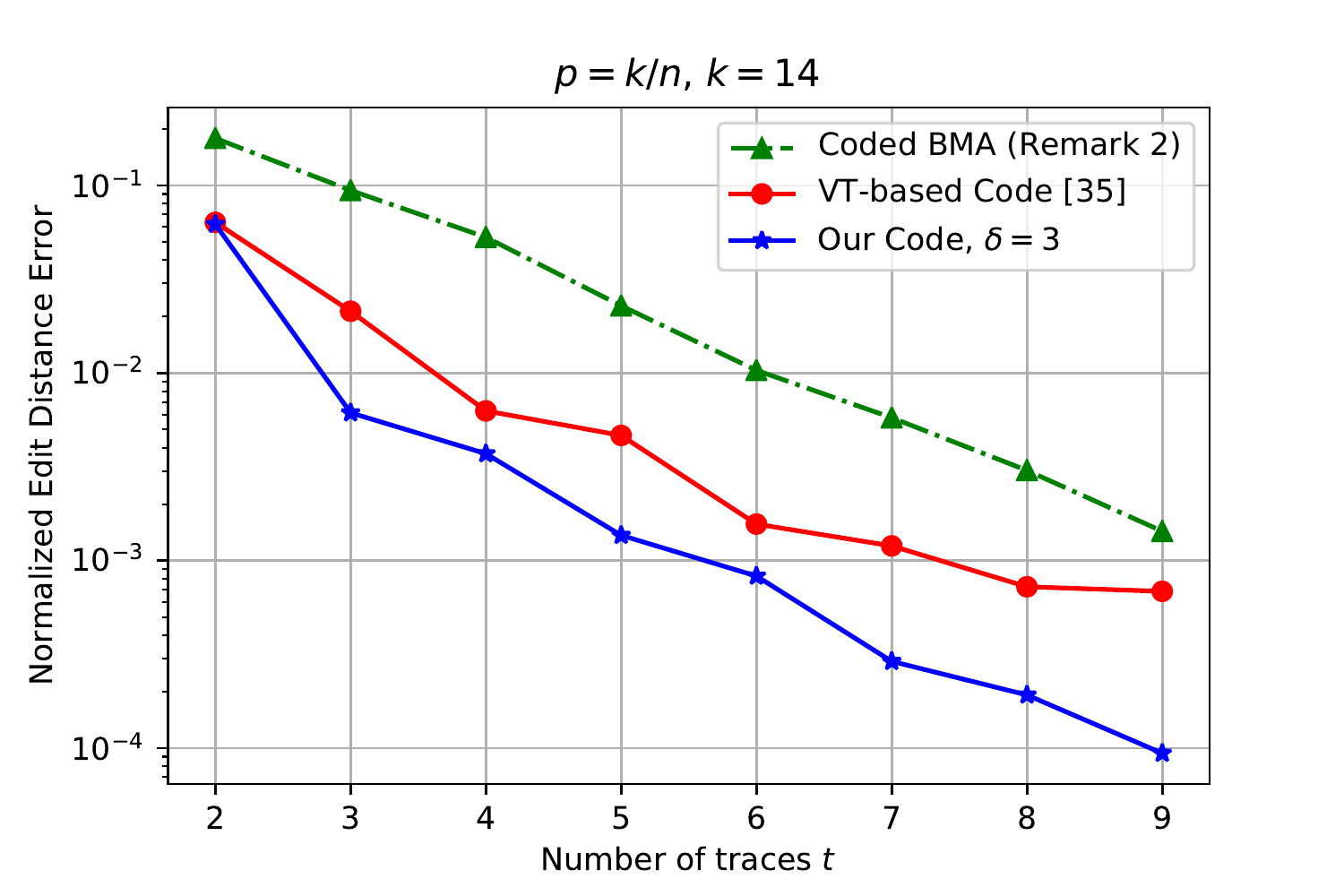}
  \caption{Normalized edit distance error as a function of the number of traces $t$ for codeword length \mbox{$n=994$} and deletion probability $p=k/n^{\alpha}$, with $k=14$ and $\alpha=1$.}\label{fig:3}
  \end{center}
\end{figure}

The simulation results are shown in Fig.~\ref{fig:1}, \ref{fig:2}, and \ref{fig:3}. The edit distance error is averaged over $10^3$ runs of simulations, and then normalized by the blocklength $n$. In each run, a codeword of length $n$ is chosen uniformly at random, and the random deletion process is applied to obtain the $t$ traces. The parameter~$\delta$ of the code $C_{\delta}'(n)$ is set to $\delta=3$ for all simulations. The results, in general, show that our code has a much lower edit distance error compared to the simple coded version of the BMA~(Remark~\ref{remBMA}). This confirms our intuition about the importance of introducing the markers and subdividing the problem of reconstructing $\mathbf{x}$ into smaller subproblems. More specifically, the subproblems are easier since they deal with a lower number of deletions; and the markers prevent potential reconstruction errors from propagating to the entire codeword resulting in a lower edit distance error. For instance, we observe in Fig.~\ref{fig:2} that for $n=3000$ the normalized edit distance error is $\approx 10^{-3}$ for our code and $\approx 2.5 \times 10^{-2}$ for the coded BMA. Note that this improvement comes at the expense of a lower code rate; however, different trade-offs can be achieved by varying the parameters $\delta$ and $\ell$. 

In Fig.~\ref{fig:3}, we also compare the performance our code to the VT-based code in~\cite{AbroshanTrace} in the regime where the average number of deletions per segment is a constant (i.e., $\alpha=1$). The code in \cite{AbroshanTrace} is constructed by concatenating $n/\ell=17$ VT codewords each of size $\ell=59$, resulting in a code with rate~\mbox{$\approx 0.9$}. The parameters of our code are set to $\delta=3$, $n=994$, and $\ell=71$, resulting in a code with rate $\approx 0.934$. The results show that our code has a lower edit distance error compared to the code in~\cite{AbroshanTrace}, for a higher code rate. 

\section{Conclusion}\label{sec:conc}
In this paper, we studied the problem of constructing codes that detect the exact number of worst-case deletions in concatenated binary strings. First, we determined the fundamental limits of the problem by deriving lower bounds on the redundancy of deletion-detecting codes. Then, we presented an optimal marker-based construction that achieves the lower bound with equality for the case of block-by-block decodable codes. The redundancy of this construction was also shown to be asymptotically optimal in the number of deletions among all families of deletion-detecting codes. We applied the marker-based construction to coded trace reconstruction and evaluated the resulting gains. Some of the open problems include finding code constructions and fundamental limits for detecting combinations of deletions and insertions. An additional interesting direction for future research is to study coded trace reconstruction for random channels that introduce both deletions and insertions.

\section*{Acknowledgment}
The author would like to thank the anonymous reviewers who reviewed this paper. Their insightful comments and valuable feedback have greatly improved the quality of this work. 

\bibliographystyle{IEEEtran}
\bibliography{Refs,Refs4}

\newpage
\appendices
\section{Proof of Theorem~\ref{thm:main1}}
First, we derive an expression of the redundancy of the code $\mathcal{C}'_{\delta}(n)$ defined in Construction~\ref{cons:full}, denoted by $r_{\mathcal{C}'}(\delta)$, in terms of the code parameter~$\delta$, where $\delta=o(\log n)$. Note that Theorem~\ref{thm1} assumes that $\ell \mid n$; however, if $\ell \nmid n$, the total number of blocks would be $\left \lceil n/\ell \right \rceil$, where the length of the last block is strictly less than $\ell$. In this case, based on Construction~\ref{cons1}, we would still need $\delta+1$ redundant bits in the last block $\x^{\left \lceil n/\ell \right \rceil}$ in order to detect the exact number of deletions in block $\x^{\left \lceil n/\ell \right \rceil-1}$. Therefore, it follows from Theorem~\ref{thm1} that the redundancy corresponding to $\mathcal{D}_{\delta-1}(\lfloor 1/p \rfloor,n)$, denoted by $r_{\mathcal{D}}(\delta)$, is 
\begin{equation}
\label{eqq1}
r_{\mathcal{D}}(\delta)=(2\delta-1)\left(\left \lceil \frac{n}{\ell} \right \rceil-1\right).
\end{equation}
The following claim gives lower and upper bounds on the number of blocks by applying standard bounds on the floor and ceiling functions. 
\begin{claim}
\label{claim1}
For $n>1$ and $p<1/2$, we have
\begin{align}
kn^{1-\alpha}-1 &\leq \left \lceil \frac{n}{\ell} \right \rceil -1 < 2kn^{1-\alpha}, \label{eqq2} \\
kn^{1-\alpha} &\leq \left \lceil \frac{n}{\ell} \right \rceil < (2k+1)n^{1-\alpha}. \label{eqq3}
\end{align}
\end{claim}
We assume for now that the claim is true and give its proof later in Appendix~\ref{sec:proofclaim}. By applying the bound in~\eqref{eqq2} to the expression of $r_{\mathcal{D}}(\delta)$ in~\eqref{eqq1} we obtain
\begin{equation}
\label{eq:bb}
(kn^{1-\alpha}-1)(2\delta-1)\leq r_{\mathcal{D}}(\delta) < 2kn^{1-\alpha}(2\delta-1).
\end{equation}
Next, we show that {\em run-length-limited} constraint in Construction~\ref{cons:full} introduces an additional redundancy that vanishes asymptotically in $n$, i.e., $r_{\mathcal{C}'}(\delta)=r_{\mathcal{D}}(\delta)+o(1)$. 

Let $X\in \mathbb{F}_2^n$ be a random variable that represents a binary sequence of length $n$ that is chosen uniformly at random. Let $L(X): \mathbb{F}_2^n\to [n]$ denote the length of the longest run in $X$. Recall that $\mathcal{L}^{n}(\sqrt{\ell})$ denotes the set of {\em run-length-limited} binary sequences of length $n$ such that the length of any run of $0$'s or $1$'s in a sequence $\mathbf{x}\in \mathcal{L}^{n}(\sqrt{\ell})$ is at most $\sqrt{\ell}$.
We define the following probabilities
\begin{align}
P_1&\triangleq P(X\in \mathcal{D}_{\delta-1}(\ell,n),X \notin \mathcal{L}^n(\sqrt{\ell})), \\
P_2&\triangleq P(X\in \mathcal{D}_{\delta-1}(\ell,n),X \in \mathcal{L}^n(\sqrt{\ell})),
\end{align}
where $\ell=\lfloor 1/p \rfloor = \lfloor n^{\alpha}/k \rfloor$. Note that 
\begin{equation}
P_2=P(X\in \mathcal{C}'_{\delta}(n))=P(X\in \mathcal{D}_{\delta-1}(\ell,n))-P_1.
\end{equation}
As a first step, we are interested in computing an upper bound on $P_1$. For the sake of brevity, we will write $\mathcal{D}_{\delta-1}$ instead of $\mathcal{D}_{\delta-1}(\ell,n)$ in what follows. 
\begin{align}
P_1 &= P(X\in \mathcal{D}_{\delta-1},L(X)>\sqrt{\ell}), \\
&= P(L(X)\geq \sqrt{\ell}+1~\big|X\in \mathcal{D}_{\delta-1}) P(X\in \mathcal{D}_{\delta-1}), \\
&\leq n\cdot \frac{1}{2^{\sqrt{\ell}+1-\delta}}\cdot \frac{1}{2^{r_{\mathcal{D}}(\delta)}}, \label{eq:19}\\
&=2^{-r_{\mathcal{D}}(\delta)}\cdot \frac{n}{2^{\sqrt{\ell}+1-\delta}}.
\end{align}
Note that \eqref{eq:19} follows from requiring that at least one of the windows of length $\sqrt{\ell}+1$ contains a run, and applying the union bound.
Therefore, 
\begin{align}
P_2 &= P(X\in \mathcal{D}_{\delta-1})-P_1, \\
&\geq 2^{-r_{\mathcal{D}}(\delta)} - 2^{-r_{\mathcal{D}}(\delta)}\cdot \frac{n}{2^{\sqrt{\ell}+1-\delta}}, \\
&= 2^{-r_{\mathcal{D}}(\delta)} \left[1- \frac{n}{2^{\sqrt{\ell}+1-\delta}}  \right]. \\
\end{align}
Hence,
\begin{equation}
|C_{\delta}'(n)|=2^n \cdot P_2\geq 2^{n-r_{\mathcal{D}}(\delta)}\left[1- \frac{n}{2^{\sqrt{\ell}+1-\delta}}  \right].
\end{equation}
Thus, we obtain the following upper bound on the redundancy
\begin{equation}
r_{\mathcal{C}'}(\delta)\leq r_{\mathcal{D}}(\delta) + \log \left[1- \frac{n}{2^{\sqrt{\ell}+1-\delta}}  \right].
\end{equation}
Since $\ell=\lfloor 1/p \rfloor = \lfloor n^{\alpha}/k \rfloor$ where $k\geq 1$ and \mbox{$\alpha\in(0.5,1]$} are constants, then we have $\ell=\Theta(n^{\alpha})$. Furthermore, since $\delta=o(\log n)$, it follows that
\begin{equation}
\lim_{n\to +\infty} \log \left[1- \frac{n}{2^{\sqrt{\ell}+1-\delta}}  \right] = \log(1)=0.
\end{equation}
Therefore,
\begin{equation}
r_{\mathcal{C}'}(\delta)\leq r_{\mathcal{D}}(\delta) + o(1).
\end{equation}
Also, $r_{\mathcal{C}'}(\delta)\geq r_{\mathcal{D}}(\delta)$ is a trivial lower bound. Therefore, we have
\begin{equation}
\label{eq:rr}
r_{\mathcal{C}'}(\delta)= r_{\mathcal{D}}(\delta) + o(1).
\end{equation}
By combining the results in~\eqref{eq:rr} and \eqref{eq:bb} we conclude the proof for the redundancy of the code given in Theorem~\ref{thm:main1}.

Next, we derive an upper bound on the probability of error, i.e., the probability that the reconstruction is unsuccessful. We define the following events. Let $D$ denote the event where the process of detecting the deletions and determining the boundaries of the blocks is successful for all the blocks and all the $t$ traces. Let $B$ denote the event where the BMA algorithm (Algorithm~\ref{algo_disjdecomp}) is successful in reconstructing \mbox{$\mathbf{x}=\langle \mathbf{x}^1,\mathbf{x}^2,\ldots,\mathbf{x}^{\lceil n/\ell \rceil}\rangle$}, i.e., successful in reconstructing all the blocks $\mathbf{x}^1,\mathbf{x}^2,\ldots,\mathbf{x}^{\lceil n/\ell \rceil}$. Let $P_e$ denote the probability of error corresponding to Construction~\ref{cons:full}. For the sake of deriving an upper bound, we assume that the reconstruction is unsuccessful if the event $D$ is not realized. Hence, we can write
\begin{align}
P_e &\leq P(D^c) + P(B^c,D) \\
&= P(D^c)+P(B^c|D)P(D) \\
&\leq P(D^c)+P(B^c|D). \label{eq:perror}
\end{align}
Recall that in Construction~\ref{cons:full} we set $\ell=\lfloor 1/p \rfloor=\lfloor n^{\alpha}/k \rfloor$, and hence $p=\Theta(1/\ell)$. In the case where the process of determining the block boundaries is successful for all traces, it follows from~\cite{Batu} that the probability of error of the BMA algorithm for each block is\footnote{The constant in the upper bound on the probability $\cO(1/\ell)$ is not explicitly given in~\cite{Batu}.} $\cO(1/{\ell})$, for $p=\Theta(1/\ell)$, a constant number of traces, and an arbitrary sequence that does not have any run of length greater than $\sqrt{\ell}$. Therefore, by applying the union bound over the $\lceil n/\ell \rceil=\lceil n/\lfloor 1/p \rfloor \rceil$ blocks we obtain
\begin{align}
P(B^c|D) &\leq \left \lceil \frac{n}{\ell} \right \rceil \cO\left( \frac{1}{\ell} \right) \\
&< (2k+1)n^{1-\alpha} \cO\left( n^{-\alpha} \right) \\
&= \cO\left(n^{1-2\alpha} \right).  \label{eq:pbma}
\end{align}
Since $\alpha\in (0.5,1]$, it is easy to see from~\eqref{eq:pbma} that $P(B^c|D)$ vanishes asymptotically in $n$. Therefore, in the case where the block boundaries are recovered successfully at the decoder using the delimiter bits, the reconstruction is successful with high probability, where the probability is over the randomness of the deletion process. 

Next, we derive an upper bound on $P(D^c)$. Let $Y_m, m\in \{1,\ldots,\lceil n/\ell \rceil\}$, be the random variable that represents the exact number of deletions in block $m$ after $\mathbf{x}$ passes through the deletion channel with deletion probability~$p$. Note that \mbox{$Y_m \sim$ Binomial$(\ell,p)$}. It follows from the construction that the boundaries of a given block are recovered successfully if and only if $Y_m<\delta$, where $\delta=o(\log n)$ is a code parameter. Let $D_j, j\in [t]$, denote the event where all the block boundaries of trace $j$ are determined successfully. Therefore, by applying the union bound over the number of blocks  $\lceil n/\ell \rceil$ blocks we obtain
\begin{align}
\label{eq:djc}
P(D_j^c)\leq \left \lceil \frac{n}{\ell} \right \rceil  P(Y_m \geq \delta).
\end{align}
Next, we upper bound $P(Y_m \geq \delta)$ by applying the following Chernoff bound on Binomial random variables.
\begin{proposition}[\hspace{-0.008cm}\cite{mitzen}]
\label{prop1}
Let $Y$ be a binomial random variable with $E[Y]=\mu$. For any $\delta>0$,
$$P(Y\geq (1+\delta)\mu)\leq \exp\left[-\mu\left((1+\delta)\ln(1+\delta)-\delta \right)\right].$$
\end{proposition}
\noindent Since $E[Y_m]=\ell p =\lfloor 1/p \rfloor p$, $p\leq 1/2$, and $\lfloor 1/p \rfloor \geq 1/p - 1$, then
\begin{equation}
E[Y_m]= \left \lfloor \frac{1}{p}\right \rfloor p \geq \left(\frac{1}{p}-1\right)p = 1-p \geq \frac{1}{2}.
\end{equation}
So by applying Proposition~\ref{prop1} to the Binomial random variable $Y_m$ we get
\begin{align}
P(Y_m\geq \delta) &\leq P(Y_m\geq \delta E[Y_m]) \\
&\leq e^{-E[Y_m]\left(\delta \ln \delta -\delta +1 \right)}\\
&\leq e^{-\frac{1}{2} \left(\delta \ln \delta - \delta +1 \right)}. \label{eq:25}
\end{align}
Therefore, by substituting~\eqref{eq:25} in \eqref{eq:djc} we get
\begin{align}
P(D_j^c) &\leq \left \lceil \frac{n}{\ell} \right \rceil  e^{-\frac{1}{2} \left(\delta \ln \delta - \delta +1 \right)} \\
&< (2k+1) n^{1-\alpha} e^{-\frac{1}{2} \left(\delta \ln \delta - \delta +1 \right)}, \label{eq:27}.
\end{align}
Let
\begin{equation}
\delta^*\left( p(n)\right) \triangleq \frac{2\ln \left( \sqrt{e}  n^{1-\alpha} p(n) \right)}{W\left(2e \ln \left( \sqrt{e} n^{1-\alpha} p(n)  \right) \right)},
\end{equation}
where $W(z)$ is the Lambert $W$ function defined as the solution of the equation $we^w=z$ with respect to the variable $w$. One can easily show that by substituting $\delta$ by $\delta^*(p(n))$ in~\eqref{eq:27} we get
\begin{equation}
\label{eqq:35}
P(D_j^c) < \frac{2k+1}{p(n)}.
\end{equation}
Since $\delta$ is an integer by construction, we set $\delta=\lceil \delta^*(p(n)) \rceil$, and the bound in~\eqref{eqq:35} still holds. Furthermore, since we consider a constant number of traces $t=\cO(1)$, then by applying the union bound over the $t$ traces we get
\begin{equation}
P(D^c) \leq \sum_{j=1}^{t} P(D_j^c) < t \cdot \frac{2k+1}{p(n)} = \cO\left(\frac{1}{p(n)}\right).
\label{eq:dc}
\end{equation}
Therefore, by substituting \eqref{eq:pbma} and \eqref{eq:dc} in \eqref{eq:perror} we obtain 
\begin{equation}
P_e < \cO\left(n^{1-2\alpha} + \frac{1}{p(n)}\right).
\end{equation}
Furthermore, the time complexity of the deletion detection algorithm is linear in $n$ (Theorem~\ref{thm1}), and the complexity of the BMA algorithm is linear in the size of input matrix $t\times \ell$ since every trace is scanned once in the reconstruction process. Hence, the complexity of the reconstruction algorithm corresponding to Construction~\ref{cons:full} is linear in $n$, which concludes the proof of Theorem~\ref{thm:main1}.

\section{Proof of Corollary~\ref{cor1}}
The result in Corollary~\ref{cor1} follows from Theorem~\ref{thm:main1} for \mbox{$\alpha \in (0.5,1)$} by substituting $p(n)=n^{2\alpha-1}$ in the expression of $\delta^*$ and also in the bound on the probability of error $P_e$; in addition to applying standard bounds on the Lambert $W$ function. For $p(n)=n^{2\alpha-1}$, we have
\begin{equation}
\delta^*=\frac{2\ln \left( \sqrt{e}n^{\alpha} \right)}{W\left(2e \ln \left( \sqrt{e} n^{\alpha}  \right) \right)},
\end{equation}
\begin{equation}
\label{eq:tb}
P_e < \cO\left(n^{1-2\alpha} + \frac{1}{n^{2\alpha-1}}\right)= \cO\left( n^{1-2\alpha} \right),
\end{equation}
where the last equality in~\eqref{eq:tb} follows from the fact that the number of traces $t$ is a constant. Since $\alpha$ is a constant, we have $$2\ln \left( \sqrt{e}n^{\alpha} \right)=\ln e + 2\alpha \ln n= \Theta(\log n).$$ Furthermore, the following bound on the Lambert $W$ function can be found in~\cite{lambert}. For every $x\geq e$,
\begin{equation}
\ln x - \ln \ln x \leq W(x)  \leq \ln x -\frac{1}{2} \ln \ln x.
\end{equation}
It follows from the bound that $W(x)=\Theta(\ln x)$, so 
\begin{equation}
W\left(2e \ln \left( \sqrt{e} n^{\alpha}  \right) \right)= W(\Theta(\log n))=\Theta(\log \log n).
\end{equation} 
\begin{sloppypar}
Therefore, we have $\delta^*=\Theta(\log n/\log \log n)$. Finally, from the bounds on $r_{\mathcal{C}'}(\delta^*)$ in Theorem~\ref{thm:main1} we can deduce that \mbox{$r_{\mathcal{C}'}=\Theta(n^{1-\alpha}\log n/ \log \log n)$}.
\end{sloppypar}

\section{Proof of Corollary~\ref{cor2}}
The result in Corollary~\ref{cor2} follows from Theorem~\ref{thm:main1} for the case where $\alpha=1$. For $\alpha=1$, we have
\begin{equation}
\delta^*=\frac{2\ln \left( \sqrt{e}p(n) \right)}{W\left(2e \ln \left( \sqrt{e} p(n)  \right) \right)},
\end{equation}
\begin{equation}
P_e < \cO\left(\frac{1}{n}+\frac{1}{p(n)}\right),
\end{equation}
and $r_{\mathcal{C}'}(\delta^*)=\Theta(\delta^*)$. Hence, it is easy to see that for \mbox{$p(n)=\omega(1)$}, the probability of error vanishes asymptotically in $n$, where $r_{\mathcal{C}'}=\omega(1)$.

\section{Proof of Claim~\ref{claim1}}
\label{sec:proofclaim}
\noindent For $x\in \mathbb{R}$, we have
\begin{align}
x-1 &<\lfloor x \rfloor \leq x \\
x &\leq \lceil x \rceil < x+1.
\end{align}
Therefore,
\begin{align}
\left \lceil \frac{n}{\ell} \right \rceil-1 &= \left \lceil \frac{n}{\lfloor 1/p \rfloor} \right \rceil -1 \\
 &< \left \lceil \frac{n}{\frac{1}{p}-1} \right \rceil - 1 \\
 &< \frac{n}{\frac{1}{p}-1} \\
 &= \frac{np}{1-p} \\
 &< 2np \label{eq59}\\ 
 &= 2kn^{1-\alpha},
\end{align}
where~\eqref{eq59} follows from the fact that $p<1/2$. Hence,
\begin{equation}
\left \lceil \frac{n}{\ell} \right \rceil <2k^{1-\alpha}+1\leq (2k+1)n^{1-\alpha},
\end{equation}
where the last inequality holds for $n\geq 1$ and is achieved with equality for $n=1$. Furthermore,
\begin{equation}
\left \lceil \frac{n}{\ell} \right \rceil = \left \lceil \frac{n}{\lfloor 1/p \rfloor} \right \rceil \geq \lceil np \rceil \geq np = kn^{1-\alpha}. 
\end{equation}

\vspace{-2cm}

\end{document}